\documentclass{aa}  

\usepackage{graphicx}
\usepackage{siunitx}
\usepackage{booktabs}
\usepackage{bm}
\usepackage{txfonts}
\usepackage{xspace}
\usepackage{comment}

\usepackage{hyperref}
\hypersetup{colorlinks=true, linkcolor=red, urlcolor=black, citecolor=blue}

\usepackage{array,multirow}
\usepackage{float}

\usepackage{gensymb}

\usepackage[dvipsnames]{xcolor}
\newcommand{\cii}{[C\,{\footnotesize II}]}
\newcommand{\hi}{\textrm{H}$\,$\textsc{i}}
\newcommand{\ha}{H$\alpha$}
\newcommand{\hb}{H$\beta$}
\newcommand{\lya}{Ly$\alpha$\xspace}
\newcommand{\oiii}{[O\,{\footnotesize III}]}
\newcommand{\mgii}{\textrm{Mg}$\,$\textsc{ii}}
\newcommand{\civ}{\textrm{C}$\,$\textsc{iv}}
\newcommand{\ovi}{\textrm{O}$\,$\textsc{vi}}
\newcommand{\nv}{\textrm{N}$\,$\textsc{v}}
\newcommand{\siiv}{\textrm{Si}$\,$\textsc{iv}}
\newcommand{\plya}{\ifmmode{P_{{\rm Ly}\alpha}}\else$P_{{\rm Ly}\alpha}$\fi}

\usepackage{orcidlink}
\newcommand{\orcid}[1]{\href{https://orcid.org/#1}{\textcolor[HTML]{A6CE39}{\aiOrcid}}}

\newcommand{\kms}        {\ifmmode{\rm \,km\,s^{-1}}\else\,km\,s$^{-1}$\xspace\fi}
\newcommand{\unitNHI}    {\ifmmode{\rm \,cm^{-2}}\else\,cm$^{-2}$\xspace\fi}  
\newcommand{\vexp}       {\relax\ifmmode {v_{\rm exp}} \else {$v_{\rm exp}$}\expandafter\xspace\fi}
\newcommand{\NHI}        {\relax\ifmmode{{ N}_{{\rm H}\textsc{i}}\xspace} \else {${ N}_{{\rm H}\textsc{i}}$}\expandafter\xspace\fi}
\newcommand{\openangle}        {\relax\ifmmode{{ \theta}_{\rm o,Wind}\xspace} \else {${ \theta}_{\rm o,Wind}$}\expandafter\xspace\fi}
\newcommand{\losangle}        {\relax\ifmmode{{ \theta}_{\rm LOS}\xspace} \else {${ \theta}_{\rm LOS}$}\expandafter\xspace\fi}
\newcommand{\sigsrc}        {\relax\ifmmode{{ \sigma}_{\rm Src}\xspace} \else {${ \sigma}_{\rm Src}$}\expandafter\xspace\fi}

\usepackage{natbib}
\bibpunct{(}{)}{;}{a}{}{,} 

\begin{document} 
\titlerunning{Constraining the geometry of the gas surrounding a typical galaxy at $z = 3.4$ with \lya polarization}
\title{Constraining the geometry of the gas surrounding \\
a typical galaxy at $z = 3.4$ with \lya polarization 
 }
\authorrunning{A. Bolamperti et al.}
\author{A. Bolamperti\inst{1,2,3} \thanks{\email{andrea.bolamperti@phd.unipd.it}}$^{\orcidlink{0000-0001-5976-9728}}$
    \and S.-J. Chang\inst{4}$^{\orcidlink{0000-0002-0112-5900}}$
    \and J. Vernet \inst{2}$^{\orcidlink{0000-0002-8639-8560}}$
    \and A. Zanella \inst{3}$^{\orcidlink{0000-0001-8600-7008}}$
    \and M. Gronke\inst{4}$^{\orcidlink{0000-0003-2491-060X}}$ 
    \and \\
    F. Arrigoni Battaia\inst{4}$^{\orcidlink{0000-0002-4770-6137}}$
    \and F. Calura\inst{5}$^{\orcidlink{0000-0002-6175-0871}}$
    \and E. Iani\inst{6}$^{\orcidlink{0000-0001-8386-3546}}$
    \and E. Vanzella\inst{5}$^{\orcidlink{0000-0002-5057-135X}}$
    }
\institute{Dipartimento di Fisica e Astronomia, Università degli Studi di Padova, Vicolo dell'Osservatorio 3, I-35122 Padova, Italy \goodbreak
        \and European Southern Observatory, Karl-Schwarzschild-Strasse 2, D-85748 Garching bei M\"unchen, Germany
        \goodbreak
        \and Istituto Nazionale di Astrofisica (INAF), Osservatorio di Padova, Vicolo dell'Osservatorio 5, I-35122 Padova, Italy \goodbreak
        \and Max-Planck-Institut f\"ur Astrophysik, Karl-Schwarzschild-Str. 1, D-85748 Garching, Germany \label{mpa}
        \goodbreak
        \and INAF -- OAS, Osservatorio di Astrofisica e Scienza dello Spazio di Bologna, via Gobetti 93/3, I-40129 Bologna, Italy
        \goodbreak
        \and Kapteyn Astronomical Institute, University of Groningen, 9700AV Groningen, The Netherlands
}
   \date{Received --; accepted --}

  \abstract
  {
  Lyman-$\alpha$ (\lya) emission is the intrinsically strongest tracer of recombining ionized hydrogen in young, star-forming galaxies, but its origin is still debated. \lya arises when emitted photons scatter in neutral hydrogen, with each scattering event changing their propagation direction and frequency. So far, observational efforts have mostly focused on the \lya surface brightness and spectral profile, which depend on the neutral hydrogen column density, geometry, kinematics, powering mechanism and on the region from which the photons are emitted. Although different processes produce similar spectra, they have different degrees of polarization, that we can use to discriminate between them and to put stringent constraints on the geometry of the galaxy and its circumgalactic medium (CGM) where \lya photons scatter, and on their emission mechanism. In this paper, we present the first deep spectropolarimetric observations of a typical clumpy star-forming galaxy at $z\sim 3.4$, strongly lensed by the cluster of galaxies Abell 2895, taken with the Polarimetric Multi Object Spectroscopy (PMOS) mode of the VLT/FORS2 instrument. We measure a \lya degree of polarization $1\sigma$ upper limit of $4.6\%$. We develop new \lya radiative transfer models assuming a biconical outflow geometry to reproduce the observations. We find that they can be explained by assuming the star-forming galaxy being embedded in a CGM with a biconical outflow geometry, with an opening angle of the wind \openangle $\sim$ 30\degree\ for line-of-sight angles \losangle $\leq 20\degree$, \openangle $\sim$ 45\degree\ for $\losangle~\leq~20\degree$, \openangle $\sim$ 60\degree\ for $\losangle~\leq~20\degree$, and \openangle $\sim$ 75\degree\ for $\losangle \leq 40\degree$, where $\losangle = 0\degree$ means observing in the direction of the outflow. We notice that the constraints from polarization are complementary to those from the spectral line profile, and the joint analysis allows us to break the degeneracies that affect them individually. This study shows the potential of including measurements of the \lya degree of polarization to constrain the symmetry of the gas surrounding typical star-forming galaxies at the cosmic noon, and paves the way to deep spatially-resolved studies of this kind, that will allow us to disentangle between different mechanisms that can originate the \lya emission.
  }


\maketitle

\section{Introduction}

The evolution of galaxies is impacted by the gas reservoirs around them, known as the circumgalactic medium \citep[CGM;][]{Tumlinson2017}. The CGM plays a central role in the exchange of gas, dust and metals between the galaxy and its surroundings. It funnels toward galaxies the gas needed to form stars, it is the site where powerful galactic outflows end up and where gas recycling happens \citep{Putman2012, fox2017, PerouxHowk2020, Veilleux2020}. The CGM presents a complex density and kinematic structure, that has been studied in the last decades thanks to multi-wavelength observations and simulations. They revealed that the CGM is ``multiphase'', as it consists of different components ranging over wide intervals of density, temperature, and ionization state \citep[e.g.,][]{Peroux2019, Wakker2012, Ford2013, Anderson2013, Suresh2017, Weng2022}. 
This complexity makes the modeling of the geometry and kinematics of the CGM challenging, in particular at high redshift, where it is usually modeled with simplified spherically or cylindrically symmetric static or expanding gas geometries (such as expanding ellipsoids) or with bipolar outflows \citep[e.g.,][]{Eide2018}, but the currently available data do not allow us to distinguish them. 

The Lyman-$\alpha$ (\lya) line is one of the best observational signatures used in such studies, in particular at high-$z$, given its dependence on the structure, ionization, and kinematics of the \hi\ gas where its photons propagate \citep{Osterbrock1962, Dijkstra2016, Gronke&Dijkstra2016}.
It is the strongest tracer of recombining ionized hydrogen in young star-forming galaxies \citep{Partridge&Peebles1967} and is ubiquitously detected at high-$z$ \citep[e.g.,][]{Zitrin2015, Vanzella2017, Caminha2023, Bunker2023, Nakane2024}, but its interpretation is still debated. In fact, differently than other lines like the \ha\ whose photons propagate undisturbed to reach us, \lya has a resonant nature, and thus a \lya photon can undergo a great number of scatterings after its emission. The number of scatterings that it experiences before being able to leave its emission site depends on the \hi\ column density, geometry, and kinematics \citep{Adams1972, Dijkstra2014}, on quantum mechanical probabilities \citep{Stenflo1980}, and on the properties of the region where it originated. For instance, centrally emitted \lya photons, e.g., created as nebular emission powered by star formation, significantly scatter before escaping, potentially giving rise to an observed spatially-extended \lya emission. However, spatially-extended \lya emission can also be produced by cooling gas \citep{Haiman2000}, gas that has been shock-heated by supernova explosions \citep{Mori2004} and galactic winds \citep{Taniguchi2000}, fluorescent radiation from an external ionizing field \citep{Hogan1987, Cantalupo2005}, or extended star formation \citep{Momose2016, Mas-Ribas2017}. These features are encoded in the intensity spectrum of the source \citep{Ahn2002, Ahn2003, Verhamme2006, Dijkstra2008, Gronke2015}, causing the broadening and shifting of its \lya line profile \citep{Neufeld1990, Dijkstra2006}. Moreover, the \lya spectrum is also affected by radiative transfer effects at the interstellar medium (ISM) and CGM scales and is dependent on the inclination and evolutionary stage of the galaxy (see \citealt{Blaizot2023}, for a study on a simulated \lya emitter galaxy at $z \sim 3-4$). The spectrum and the \lya surface brightness profile, that reveals the spatial distribution of the \lya emission and the diffusion process of \lya photons, are the most frequently used observables embraced to investigate the nature of the \lya emission. But, thanks to its resonant nature, we can additionally leverage on the \lya degree and direction of linear polarization, typically represented by the Stokes parameters $Q$ and $U$. 

The \lya degree of polarization increases if photons are scattered in a preferential direction, and the resulting value mainly depends on two fundamental factors: the production mechanism and site where the \lya photons are created, and the geometry of the gas where they scatter before escaping, being in particular sensitive to the isotropy and homogeneity of the emission and gas distribution \citep[e.g.,][]{Lee1998, Ahn2002, Eide2018}. 
Theoretical studies showed that different models may present similar spectra, but different degrees of polarization of the \lya line \citep[e.g.,][]{Dijkstra2008,Gronke2015, Eide2018}. 

Due to the observational difficulty of measuring the \lya degree of polarization of distant sources and the limited number of available (spectro-) polarimeters intended for extragalactic use \citep{Hayes&Scarlata2011, Beck2016}, only a few studies that map the polarization of \lya are currently present, and mostly with the narrow-band imaging mode. Moreover, they all target bright ($L_\mathrm{Ly\alpha} > 10^{43}$~erg\,s$^{-1}$) and extended (up to $\sim 150$~kpc) \lya emissions at $z \sim 2-3$ in dense environments, with no clear central powering sources (the so-called ``\lya blobs''; LAB) or around extreme environments, such as high-$z$ quasars, overdense regions (clusters or protoclusters) and AGN or radio galaxies. The interpretation of the results remains questionable due to a lack of general consensus. 
\citet{Koratkar1995} made use of the HST Faint Object Spectrograph (FOS) spectropolarimeter to observe three QSOs at $z=0.5-1.6$ including PG~1630+377, which presents an increase of the degree of polarization up to $\sim 20 \%$ blueward the Lyman break, and a \lya line polarization of $(7.3 \pm 1.6)\%$, that can be explained with the presence of two sources, one completely obscured and producing the scattered polarized signal, and a secondary unpolarized redder source. \citet{Vernet2001vlt} presented low resolution VLT/FORS1 spectropolarimetry of the submillimeter-selected galaxy SMM J02399$-$0136, finding a \lya degree of polarization of $2.1^{+0.9}_{-0.5}\%$. \citet{Vernet2001keck} measured \lya degrees of polarization $<2\%$ in nine radio galaxies at $z \sim 2.5$ with Keck II/Low Resolution Imaging Spectrometer (LRIS).
\citet{Prescott2011} observed the LAB LABd05, containing an obscured AGN at $z=2.656$. Due to the coarse spatial resolution, they could only put an upper limit of $\approx 5\%$ on the \lya degree of polarization within an aperture of 65~kpc. With deeper data, \citet{Kim2020} found a consistent polarization fraction of $(6.2 \pm 0.9) \, \%$ within the same aperture, but could also detect a spatially resolved polarization varying from 5\% at the \lya peak to 20\% 45~kpc away, consistent with \lya photons not being scattered in the central region between the AGN and the \lya peak, but only in the outer gas surrounding the nebula. Similarly, \citet{Humphrey2013} measured a low ($<5\%$) polarization fraction in the center of the LAB hosting the TXS 0211$-$122 radio galaxy at $z = 2.34$, increasing to $(16.4 \pm 4.6) \, \%$ in the eastern section. \citet{You2017} found no polarized signal in the center of the LAB B3, which surrounds a radio-loud AGN at $z = 3.09$, and a degree of polarization of 3\% (17\%) at 10 (25)~kpc, with an asymmetric distribution. 
\citet{Hayes2011} observed LAB1, in the SSA22 protocluster at $z = 3.09$, and did not detect polarized signal in the center, but a 20\%-polarized ring at approximately 45~kpc. The results were confirmed by using spectropolarimetric observations by \citet{Beck2016}, who found increasing polarization towards the wings of the \lya spectral profile, which can be explained by the presence of outflows. Making use of Integral Field Spectroscopy observations for this system, \cite{Herenz2020} found that regions with a larger degree of polarization also have high velocity shifts and narrow line profiles, and associated these evidences with \lya scattering from a central source. Finally, \citet{North2024} did not detect polarized signal in the LAB hosting the radio-quiet quasar SDSS J1240+145, at $z = 3.11$.

Additionally, state-of-the-art models and simulations are needed to interpret the observations. In the last few years, significative steps forward have been done in such studies, with the implementation of \lya polarization in advanced \lya radiative transfer codes \citep[e.g.,][]{Ahn2000, Dijkstra2008, Ahn2015, Chang17, Eide2018, Seon22, Chang2023, Chang2024}, that allow us to predict the polarization behavior in different geometries (spherically symmetric like the expanding shell, and non spherically symmetric like the expanding ellipsoids or biconical outflows) and with different physical properties (density and clumpiness of the gas) and kinematics. 

In this paper we investigate the \lya origin and the geometry of the scattering CGM through the first study of the degree of polarization of the \lya emission line for a typical clumpy, star-forming galaxy \citep[Abell~2895a in][]{Livermore2015} at $z \sim 3.4$, strongly lensed by the cluster of galaxies Abell 2895 (A2895) into three multiple images. 
Thanks to the image multiplicity and the lensing magnification, we are able to study the properties of this source in great detail using the available multi-wavelength dataset, which includes ancillary observations taken with the Hubble Space Telescope (HST), the Multi Unit Spectroscopic Explorer (MUSE), the Enhanced Resolution Imager and Spectrograph (ERIS), and the Spectrograph for INtegral Field Observations in the Near Infrared (SINFONI) at the Very Large Telescope (VLT), the Atacama Large Millimeter/submillimeter Array (ALMA), and new observations taken with VLT/FOcal Reducer/low dispersion Spectrograph 2 (FORS2) instrument, with its Polarimetric Multi Object Spectroscopy (PMOS) mode.

The paper is organized as follows. 
In Section~\ref{sec:data}, we present the data available for this system, summarizing the ancillary archival data and then focusing on the new FORS2 PMOS observations. We also describe the pipeline and the step followed in the data reduction.
In Section~\ref{sec:analysis3}, we describe the 1D spectra extraction, the dilution correction, the measurement of the Stokes parameters and polarization fraction as a function of the wavelength, and the assumed bins in wavelength. In Section~\ref{sec:radiative_transfer} we expose the radiative transfer models we developed to interpret the results, and we compare observations and models in Section~\ref{sec:comparison_obsmod}.
In Section~\ref{sec:discussion} we discuss our results and the future perspectives.
In Section~\ref{sec:summary}, we summarize our results.

Throughout this paper, we adopt a flat $\Lambda$CDM cosmology with $\Omega_\Lambda = 0.7$, $\Omega_{\rm m} = 0.3$, and $H_0 = 70$~km s$^{-1}$ Mpc$^{-1}$. We report all the measurements corrected for lensing effects.

\section{Data}
\label{sec:data}
\subsection{Our target galaxy and its ancillary data}
Our target, Abell~2895a, is a clumpy star-forming (${\rm SFR} = 10 \pm 0.3$~M$_\odot$\,yr$^{-1}$) galaxy at $z\sim 3.4$, characterized by an extended \lya emission with a total flux of $(1.41 \pm 0.04) \times 10^{-17}$~erg\,s$^{-1}$\,cm$^{-2}$, offset by $1.2 \pm 0.2$~kpc with respect to the UV continuum \citep{Iani2021, Zanella2024}.
Abell~2895a is strongly lensed by the galaxy cluster A2895 into three multiple images with coordinates (M1; M2; M3) (RA, dec) = (01:18:11.19, $-$26:58:04.4; 01:18:10.89, $-$26:58:07.5; 01:18:10.57 $-$26:58:20.5) \citep{Livermore2015, Iani2021}. They are located in the inner region of the A2895 cluster, angularly close to the brightest cluster galaxy (BCG; at $z = 0.227$). Through this work, we will focus on the two most magnified images, M1 and M2, whose average lensing magnifications are $\mu = 5.5 \pm 0.7$ and $\mu = 4.5 \pm 0.3$, respectively \citep{Iani2021}.

Abell~2895a has a redshift of $z_\mathrm{opt} = 3.39535 \pm 0.00025$, estimated from optical emission lines \citep{Iani2021}, which is consistent with the $z_\mathrm{\cii} = 3.39548 \pm 0.00007$ estimated from the \cii\, far-infrared line \citep{Zanella2024}. It presents a clumpy morphology in the UV continuum and in \cii\, emission line observations. At least four star-forming clumps within a diffuse emission are detected in the UV continuum, one of which is also observed in the \cii\ data. Two additional clumps, without a detected UV counterpart, are identified in \cii, whereas no dust continuum is detected down to $F_\mathrm{cont} < 34 \, \mu$Jy \citep{Zanella2024}, in agreement with the measured blue UV-continuum $\beta$ slope of $-2.53 \pm 0.15$ and low reddening of $E(B-V) < 0.16$~mag \citep{Iani2021}.

\begin{figure*}[!h]
    \centering
    \includegraphics[width=0.97\textwidth]{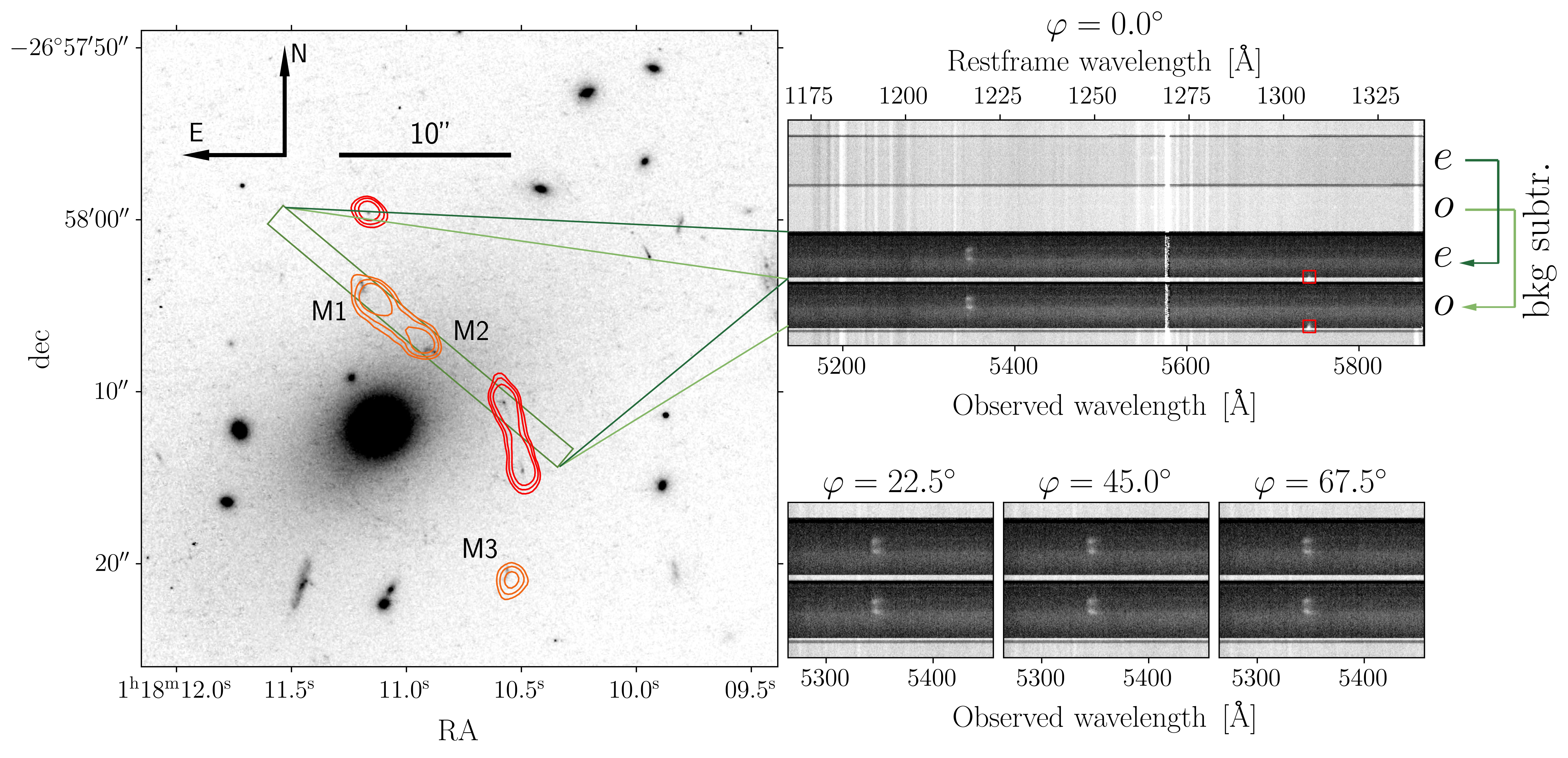}
    \caption{Left: HST F606W image of the inner part of A2895, where the three multiple images of Abell~2895a (M1, M2, and M3) appear. It represents the rest-frame UV at the redshift of Abell~2895a, $z \sim 3.4$. In orange, we show the 2, 3, and 5$\sigma$ contours of the \lya emission, detected with MUSE. The green box represents the FORS2 $1.4\arcsec \times 22$\arcsec adopted slit, that includes M1 and M2, and one image of another source at $z \sim 3.7$ \citep{Iani2023}, whose \lya 2, 3, and 5$\sigma$ contours are shown in red. Thanks to the PMOS mode, the signal included in the slit is split in an ordinary ($o$) and extraordinary ($e$) ray, with orthogonal polarizations, shown with different tones of green. Right: stacked 2D spectrum with $\varphi = 0.0\degree$ obtained after calibration, cosmic ray rejection, and sky subtraction. In particular, the sky is subtracted only in the relevant slits which contain Abell~2895a, and appear darker. Each slit has a spatial (vertical) width of 22\arcsec, and the sky level is estimated and then subtracted from the closest ordinary and extraordinary slits (see the $e$ and $o$ labels on the right), respectively, as shown by the green arrows on the right. In the slits including the target, we detect the extended \lya emission associated with M1 and M2, at approximately 5340 \AA, the continuum from the nearby BCG, in the bottom part, and the \lya emission of the source at $z\sim 3.7$ (red squares, close to the edge of the slitlets). We show the 2D spectrum taken with $\varphi = 0.0\degree$ on top, and smaller cutouts around the \lya line for $\varphi = 22.5\degree$, 45.0\degree\, and 67.5\degree\ on the bottom.}
    \label{fig:system_slit}
\end{figure*}
Abell~2895a offers a suite of ancillary data, presented in \citet{Iani2021} and \citet{Zanella2024}. 
The rest-frame UV imaging was observed with HST (SNAP program 10881, PI:~G.~Smith) with a $\mathrm{FWHM} \sim 0.13\arcsec$ resolution in the F606W filter. The \hb\ and \oiii\ lines have been studied with VLT/SINFONI spectra \citep{Livermore2015}, and new VLT/ERIS observations (program IDs 110.2576, 112.25HA, 114.273Y, PI:~A.~Zanella), aimed at spatially-resolving them, are ongoing.
The \lya emission was targeted by VLT/MUSE observations (program IDs 60.A-9195(A), 0102.B-0741(A), PI:~A.~Zanella) in the Adaptive Optics (AO) Wide Field Mode (WFM), with a resolution of $\mathrm{FWHM} \sim 0.4\arcsec$.
It appears spatially offset with respect to the clumps visible in the UV continuum and the optical emission lines. The HST rest-frame UV and the \lya contours are shown in Fig.~\ref{fig:system_slit}. The \lya line also presents an asymmetric profile, that is redshifted with a relative velocity $\Delta v=403 \pm 4$~\kms with respect to the systemic redshift \citep{Iani2021}. 
Finally, ALMA Band 8 observed Abell~2895a (program ID 2019.1.01676.S, PI:~E.~Iani) to detect the \cii\ emission line and the underlying continuum \citep{Zanella2024}. The beam size (assuming a natural weighting) is $\mathrm{FWHM}=0.31\arcsec \times 0.26 \arcsec$.

A2895 also benefits from a robust strong lensing model, introduced by \citet{Iani2021}. The 2D-projected total mass distribution of the cluster is modeled as a combination of an extended cluster-scale halo and multiple galaxy-scale double pseudo-isothermal elliptical components \citep[dPIE,][]{Eliasdottir2007}, whose centers and shapes are constrained by the respective surface brightness centroids, ellipticities, and position angles from the HST F606W image. The cluster members are selected through the color–magnitude diagram method (e.g. \citealt{Richard2014}), and the total mass associated to each member is computed from its luminosity, through the Faber–Jackson relation for elliptical galaxies \citep{Faber1976}, as is usual in strong lensing modeling on cluster scales \citep[e.g.,~][]{Caminha2023, Bergamini2023}. The model is constrained by using the location of the multiple images of Abell~2895a and those of another triply imaged system with spectroscopic redshift $z = 3.721$ \citep{Livermore2015, Iani2023}, shown in Fig.~\ref{fig:system_slit}. The best-fit model reproduces the location of the multiple images with a root-mean-square (rms) displacement of 0.09\arcsec\ between the observed and modeled multiple images positions. The convergence maps derived from this strong lensing model are exploited to estimate the magnification factors in the locations of M1 and M2, that we use to derive the magnification-corrected quantities adopted in the following.

\subsection{FORS2 PMOS observations and data reduction}
Abell2895a was observed with VLT/FORS2 \citep{Appenzeller1998} between September 2021 and August 2022 (program ID 108.2260, PI:~A.~Zanella), for a total of 18.1 hours in PMOS mode. The polarization optics are composed of a superachromatic half-wave plate mosaic followed by a Wollaston prism, that separates the light into two beams with orthogonal polarization (the "ordinary" ($o$) and "extraordinary" ($e$) rays). Half of the MOS mask slitlets in front of the polarization optics are fully closed to avoid the overlap of the $o$ and the $e$ beams, leaving eight 22\arcsec \, high slitlets for science targets. 

Observations were executed with seeing $<0.9$\arcsec, clear sky conditions, fraction of lunar illumination $< 0.4$, and airmass $\lesssim 1.6$. The run was divided into twelve sets of four 1200~s exposures with the half wave plate position angles ($\varphi$) set successively to 0\degree, 22.5\degree, 45\degree, and 67.5\degree. The target acquisition was performed, with a typical precision of $<0.1\arcsec$, through a blind offset from a bright star, distant $\sim 40$\arcsec\ from the target.
We used the MIT red CCD together with the 1400V grism and a slit width of 1.4\arcsec\ for all observations providing an effective spectral resolution of about 3.6~{\AA} FWHM covering the wavelength range from 4560 to 5860 \AA. The slit was oriented at 40\degree\ North to East so that both M1 and M2 fit in a single slitlet (see Fig.~\ref{fig:system_slit}), which also includes one multiple image of another source at $z \sim 3.7$ \citep{Iani2023}. However, this galaxy is too close to the edge of the slit and the slit losses are too important to analyze the polarization of this second target too, that is not considered in the following analysis.

We reduced the data with the standard FORS2 PMOS pipeline v5.14, making use of the ESO Recipe Execution Tool \citep[\texttt{EsoRex},][]{esorex} pipeline. We reduced separately the observations of each of the twelve OBs and combined them as the last step. We focused on the data taken with the CHIP1 (Norma) CCD, which contains the spectrum of Abell~2895a in the bottom part.
We ran the calibration recipe to correct all the raw exposures using the associated \texttt{BIAS} frames, identify the slitlets limits, the dispersion relation, and the spatial distortion, and correct for the \texttt{FLAT} fields. These products were used as inputs in the science recipe, that produces as output wavelength-calibrated optical distortion-corrected 2D spectra. Given the faintness of our target in the single OB, we disabled the sky subtraction automatically performed by the pipeline, as it may affect the resulting signal-to-noise ratio ($S/N$) of the target. 
We detected and rejected the cosmic ray traces with the \texttt{Astro-SCRAPPY} \citep{McCully2018zndo} Python package, based on L.A.Cosmic \citep{Dokkum2001}. To properly subtract the sky background contribution for the $o$ and $e$ beams, we estimated the median flux at each wavelength in the two respective closest slits, located $22\arcsec$ apart (Fig.~\ref{fig:system_slit}, right panel, arrows on the right). 
We did not use the slits containing the target themselves to estimate the sky as they are dominated by the emission of M1, M2, and the BCG contribution. We checked that, after subtracting the median sky flux from the target slits, the residuals did not show systematics or gradients. Three OBs presented a strongly polarized background contamination, likely due to the presence of the moon, and thus with a not robust sky subtraction. We decided to exclude them so as not to bias our analysis. These OBs were the only ones taken with the moon above the horizon, and had the lowest angular separation between the target and the Moon ($\sim$ 90\degree\ instead of the 140\degree\ - 150\degree\ of the rest of the OBs).
We stacked the remaining nine OBs by spatially matching the position of the \lya peak. We computed, for each OB, a profile in the spatial $y$-direction around the \lya line and identified the peaks of the two \lya glows, relative to M1 and M2. We noticed that, in the selected OBs, the peaks do not show significant shifts ($\lesssim 0.5 \arcsec$, much smaller than the extraction aperture used for 1D spectra in the following), and thus we directly stacked them, obtaining four 2D spectra, one for each of the four $\varphi$ values (shown in Fig.~\ref{fig:system_slit}). We measured the 2D variance spectra by converting the observed spectra from ADU/s to total counts, and propagating the uncertainties associated to the object, the sky, and the readout noise assuming Poissonian statistics. 

The four resulting 2D spectra of the $o$ and $e$ channels including Abell~2895a consist of 22\arcsec\ slits, with a spatial pixel sampling of $\sim \,$0.25\arcsec\ pix$^{-1}$ and spanning in wavelength from 4560 \AA\ to 5860 \AA, with an effective spectral resolution of about 3.6~{\AA} FWHM, and with a 0.64 \AA\ pix$^{-1}$ dispersion.

\section{Analysis}
\label{sec:analysis3}
\begin{figure*}
    \centering
    \includegraphics[width=0.8\textwidth]{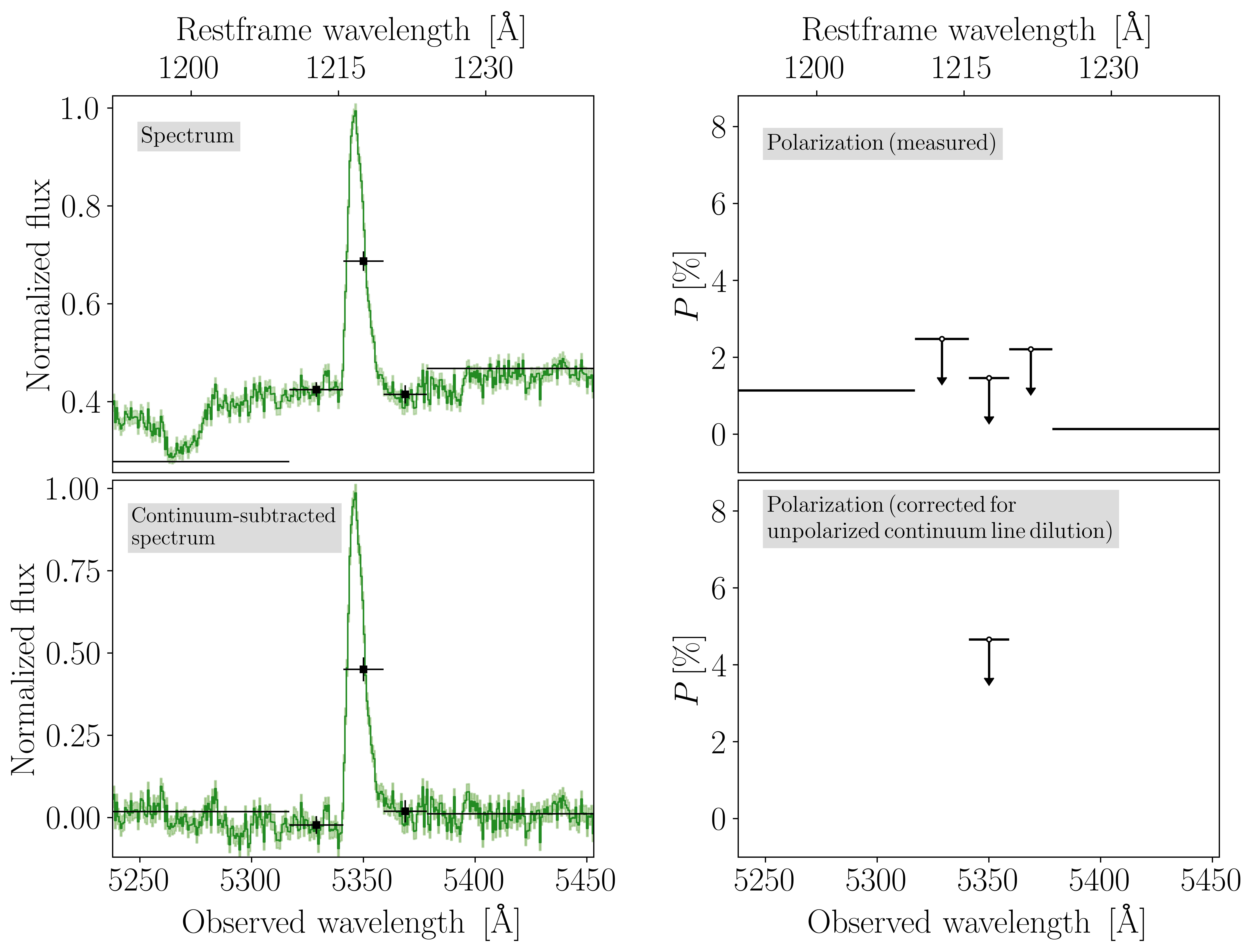}
    \caption{Left panels: total intensity spectra (green solid line) and 1$\sigma$ uncertainties (light green shaded region) in the spectral region around the \lya line. The top (bottom) panel shows the normalized spectrum before (after) the subtraction of the BCG contribution. The black filled squares, with 1$\sigma$ uncertainties, represent the binned data. Right panels: polarization ($P$) measurements obtained before (top) and after (bottom) the dilution correction described in Eq.~\ref{eq:dilution}. The black open circles represent the $1\sigma$ upper limits obtained by applying the correction of \citet{Simmons1985}. In the bottom right panel, only one datapoint is visible, as the others are outside the plotted range (upper limits for $P \sim 80\%-95\%$).}
    \label{fig:pre_post_dil_corr}
\end{figure*}
\subsection{Intensity spectrum of Abell~2895a and dilution correction}
To maximize the $S/N$, we extracted the 1D spectra by summing, for each wavelength, the signal included in an aperture of $7.5\arcsec$, designed to encompass both the M1 and M2 \lya emission. After ray-tracing it to the source plane, this aperture corresponds to a physical size of approximately 10~kpc at $z=3.4$.
The eight resulting 1D spectra, with intensities $I^{o,e}_{\varphi}(\lambda)$, relative to the $o$ and $e$ channels with $\varphi = 0.0\degree, 22.5\degree, 45.0\degree$, and $67.5\degree$, showed a significant contribution from the flux of the angularly close BCG. This signal is not polarized and it does not directly affect the polarization measurements of the \lya line of Abell~2895a, but dilutes it. Thus, the maximum polarization fraction we could measure was limited by the dilution factor, $f_{\rm d}$. 
We firstly estimated the total intensity spectrum $I(\lambda)$ within the extraction aperture, that is equivalent to the total intensity Stokes parameter spectrum, as the sum of the eight 1D spectra. Then, we evaluated the total intensity spectrum associated with the BCG, $I_\mathrm{BCG}(\lambda)$, by applying the same procedure to eight spectra extracted within $4\arcsec$-apertures centered on the bottom part of each slit, where there is no contribution from the \lya blobs. We normalized the $I(\lambda)$ and $I_\mathrm{BCG}(\lambda)$ spectra, and smoothed $I_\mathrm{BCG}(\lambda)$ to mitigate noise features. The $I(\lambda)$ spectrum, before and after subtracting $I_\mathrm{BCG}(\lambda)$, is shown in the left panels of Fig.~\ref{fig:pre_post_dil_corr}. We derived the dilution factor, defined as the fraction between the total intensity spectrum from the underlying continuum, mainly due to the BCG, and that of Abell~2895a only, as 
\begin{equation}
    \label{eq:dilution}
    f_{\rm d}(\lambda) = \frac{I_\mathrm{BCG}(\lambda)}{I(\lambda)-I_\mathrm{BCG}(\lambda)} \, .
\end{equation}
We measured $f_{\rm d}(\lambda)$ values ranging from approximately 2 at the peak of the \lya to 10 in the tails. 

\subsection{The reduced Stokes parameters $q$, $u$, and $P$ spectra and variance spectra}
We combined the measured $I^{o}_{\varphi}(\lambda)$ and $I^{e}_{\varphi}(\lambda)$ intensities to measure the reduced Stokes parameters $q$ and $u$\footnote{
$q(\lambda)$ and $u(\lambda)$ are measured as
$$ q(\lambda) = \frac{R_\mathrm{Q}(\lambda)-1}{R_\mathrm{Q}(\lambda)+1} \, , \quad {\rm where} \,\, R_\mathrm{Q}^2(\lambda) = \frac{I_{0.0\degree}^o (\lambda) / I_{0.0\degree}^e (\lambda)}{I_{45.0\degree}^o(\lambda) / I_{45.0\degree}^e(\lambda)} \, , $$   
    
$$ u(\lambda) = \frac{R_\mathrm{U}(\lambda)-1}{R_\mathrm{U}(\lambda)+1} \, , \quad {\rm where} \,\, R_\mathrm{U}^2(\lambda) = \frac{I_{22.5\degree}^o(\lambda) / I_{22.5\degree}^e(\lambda)}{I_{67.5\degree}^o(\lambda) / I_{67.5\degree}^e(\lambda)} \, .$$
}
as a function of the wavelength. Combining observations taken with four different position angles is crucial to correctly handle the different gain factors in the $o$ and $e$ channels, and to extract reliable $q(\lambda)$ and $u(\lambda)$ spectra \citep{Cohen1997}. The degree of polarization is then measured as 
\begin{equation}
\label{eq:pol}
    P(\lambda)=\sqrt{q^2(\lambda)+u^2(\lambda)} \, ,
\end{equation}
and the uncertainties on $q(\lambda)$, $u(\lambda)$, and $P(\lambda)$ are estimated through the propagation of the uncertainties associated with each of the eight $I_\varphi^{o}(\lambda)$ and $I_\varphi^{e}(\lambda)$ 1D intensity spectra. In principle, $q(\lambda)$ and $u(\lambda)$ allow us to measure also the polarization angle but, given the low $S/N$ regime, we could not obtain significant measurements, and will not consider the polarization angle in the following.

The degree of polarization measured through Eq.~\ref{eq:pol} is, by definition, a positive quantity. In low $S/N$ regimes, the uncertainties on $q(\lambda)$ and $u(\lambda)$ lead to an increase of the biased measured value of $P(\lambda)$, that will differ from the true unbiased value $P_0(\lambda)$. We applied the correction by \citet{Simmons1985}, that proposed, for different $S/N$ regimes, four different possible methods to estimate $P_0(\lambda)$ (the average estimator from \citealp{Serkowski1958}, the \citealp{Wardle1974} estimator, the maximum likelihood estimator, and the median estimator) and their uncertainties. In high $S/N$ regimes all the four methods predict consistent results, while they are particularly effective in estimating $P_0(\lambda)$ in low $S/N$ regimes, like those presented in this work. In this regime, the quality of the measured $q(\lambda)$, $u(\lambda)$, and $P(\lambda)$ can be moreover enhanced by binning the spectra in a number of bins that depends on the target $S/N$ and on the spectral resolution needed for a proper interpretation of the data. 
We adopted two different approaches: we included the entire \lya line in a single bin, integrating from 1215.5~\AA\ to 1219.6~\AA, or we divided it into three bins, one including the blue tail (1215.5-1216.3~\AA), a central one including the peak (1216.3-1217.9~\AA) and one including the red tail (1217.9-1219.6~\AA). 
We also included two bins to sample the continuum blueward (one narrower and closer to the \lya, from 1210.0~\AA\ to 1215.5~\AA, and one broader, from 1037~\AA\ to 1210~\AA) and redward (similarly, from 1219.6~\AA\ to 1224.0~\AA, and from 1224~\AA\ to 1260~\AA) the \lya emission. These are partially visible in Fig.~\ref{fig:pre_post_dil_corr} as the horizontal black lines.
In the following, we will refer to the case with a single bin for the \lya, while the results for the other case are shown in Appendix~\ref{app_a}, as they are equivalent and bring to the same conclusions.

We applied Eq.~\ref{eq:pol}, and measured low $P$ values, consistent with zero. After the positive bias correction and the binning, we obtained observational upper limits on $P_0$ in the considered bins. We multiplied these upper limits by the dilution factor $f_{\rm d} (\lambda)$, correspondingly binned within the same chosen spectral windows (e.g., $f_{\rm d} \approx 3$ in the \lya bin). The measured polarization fractions before and after the dilution correction are shown in the right panels of Fig.~\ref{fig:pre_post_dil_corr}. The $P_0$ values revealed that we can put tighter constraints on the polarization fraction at the peak of the \lya line, where the $S/N$ is at its maximum, and increasingly shallower ones moving towards the tails. Far from the line, in the continuum where $I(\lambda) \approx I_\mathrm{BCG}(\lambda)$, it is not possible to put informative constraints. 
After the dilution correction, we measure for the bin including the \lya $1\sigma$, $2\sigma$ and $3\sigma$ unbiased upper limits on the degree of polarization, \plya, of $4.6\%$, $5.8\%$, $6.5\%$, respectively. Due to the large dilution factor of approximately 35 (100), the degree of polarization is barely (not) constrained for the narrow (broad) continua bins. 

We checked whether the \plya\ measurements could be affected by our choice of the aperture adopted to extract the spectra, including two multiple images of Abell~2895a. In fact, the strong lensing critical lines at the redshift of Abell~2895a pass between the two images, that result fairly mirrored (as can be seen from the UV morphology and the \lya contours on the left panel of Fig.~\ref{fig:system_slit}). We extracted the spectra separately from the the extended \lya of M1 and M2, visible on the right panel of Fig.~\ref{fig:system_slit}, with apertures of 4\arcsec\ centered on the peaks (against the 7.5\arcsec aperture that includes both peaks). By adopting the same \lya bin and corrections, we obtained consistent $1\sigma$ \plya\ upper limits of 7.4\% and 5.8\%, that are less stringent due to the lower $S/N$. 

Both in the case of including the \lya emission from the M1 and M2 multiple images together or separately, the adopted extraction apertures globally contain the \lya emission from the entire galaxy. It has been observed that this approach can lower the measured degree of polarization, as the result of the cancellation of opposite contributions from different sides of the emission \citep[e.g.,][]{Humphrey2013, You2017}. In order to check whether this effect could have affected the low polarization we measured, we extracted the spectra from two regions, including separately the two different spatial halves of the \lya emission. The results of this test are described in detail in Appendix~\ref{app_b}. We found, for the two different regions, upper limits on \plya\ of 5.1\% and 7.5\% at the 1$\sigma$ level, consistent with those found in our reference case. The low $S/N$ did not allow us to include the polarization spatial information in our analysis and comparison with the models, as we will discuss in Section~\ref{subsec:future}.

\section{Radiative transfer models for \lya}\label{sec:radiative_transfer}
\begin{figure}
    \centering
    \includegraphics[width=\columnwidth]{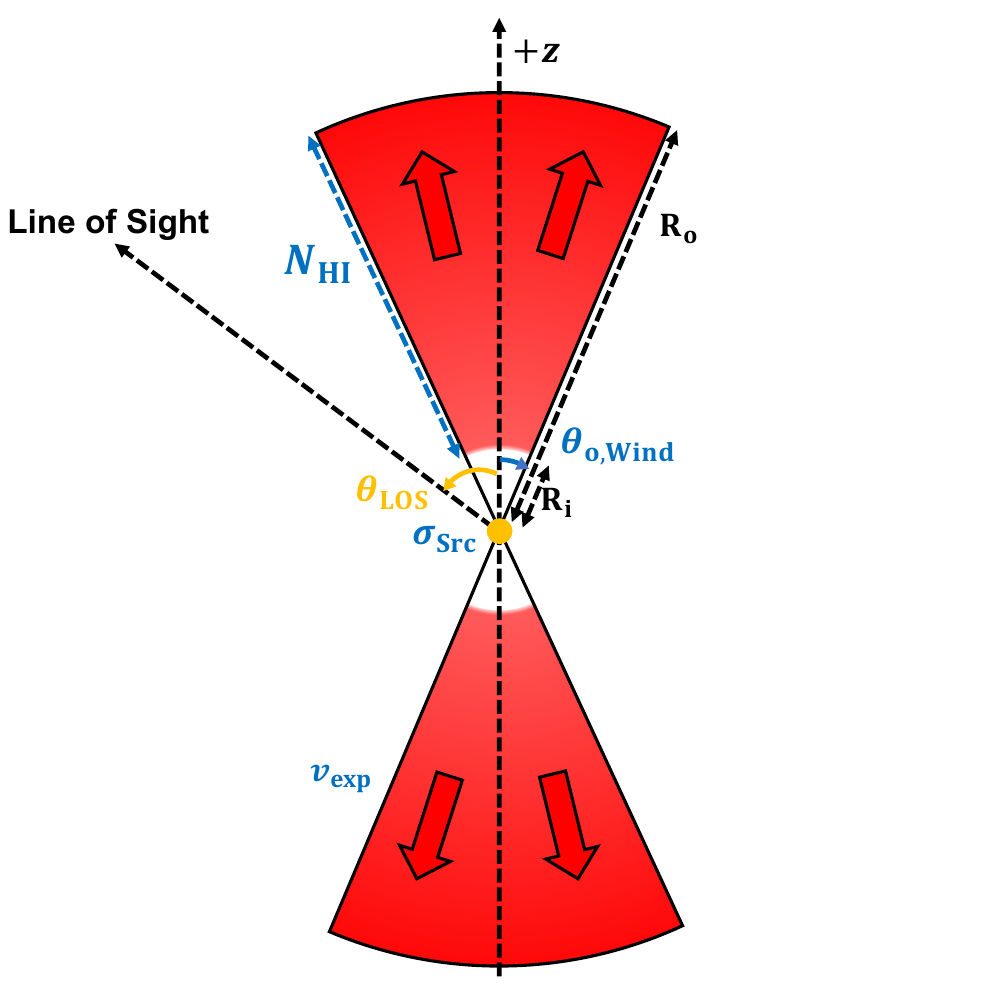}
    \caption{Schematic illustration of the wind model composed of a central point source (orange) and a bipolar outflow with the radius $R_o$ (red). The central source emits \lya photons, following a Gaussian profile with a width \sigsrc. The bipolar outflow is characterized by the \hi\ column density \NHI, the opening angle \openangle (increasing from the $+z$-axis, as the blue solid arrow), and the expansion velocity \vexp parameters. The inner radius of the outflow $R_i$ is fixed at $0.1 R_o$. As the wind model is symmetric about the $z$-axis, the line of sight angle \losangle is the angle from the $+z$-axis following the orange arrow, and thus with $\losangle=0\degree$ meaning observing in the direction of the outflow, and $\losangle =90\degree$ representing the equatorial view.
    }                        
    \label{fig:geometry}
\end{figure}
\begin{figure*}
    \centering
    \includegraphics[width=\textwidth]{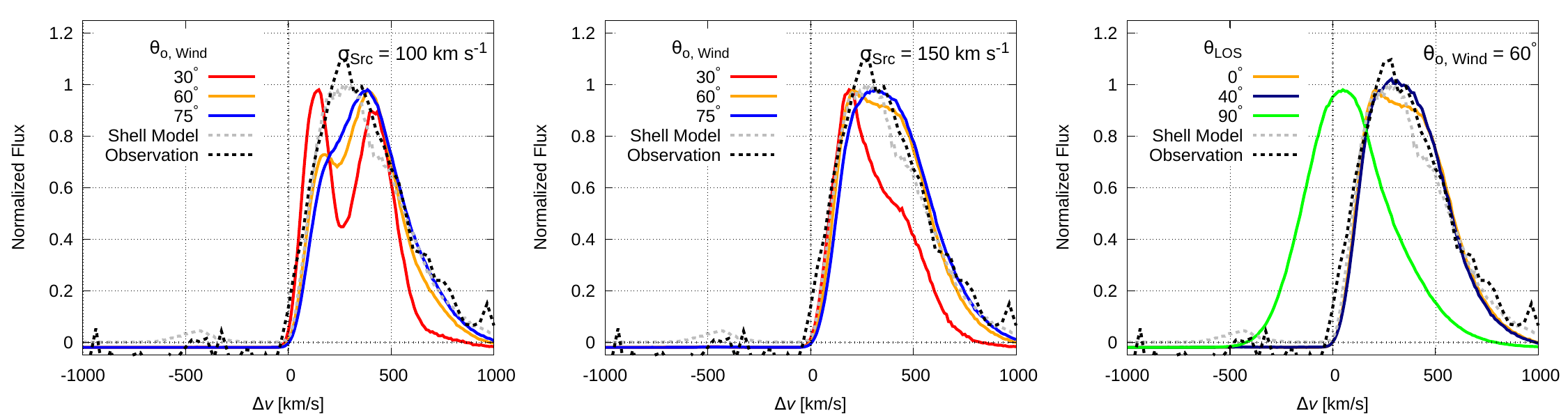}
    \caption{Comparisons between observed and simulated \lya spectra. The simulated spectra of the wind models are normalized by setting the same peak height as the spectrum of the previous fit with the shell model. The black dashed line is the observed \lya spectrum, which is continuum subtracted. The grey dashed line is the simulated \lya spectrum of the shell model from \cite{Iani2021}. In the left and middle panels line colors represent, for \sigsrc equal to, respectively, 100 \kms and 150 \kms, \openangle = 30\degree\ (red), 60\degree\ (orange), and 75\degree\ (blue), with \losangle fixed to 0\degree. In the right panel, \openangle is fixed to 60\degree, and line colors represent \losangle = 0\degree\ (orange), 40\degree\ (blue), and 90\degree\ (green).
    }                        
    \label{fig:spec_RT}
\end{figure*}

\cite{Iani2021} modeled the spectral profile of \lya through \lya radiative transfer modeling \citep{Dijkstra2019}, using the fitting pipeline of \citet{Gronke2015}, adopting the shell model \citep[e.g.,][]{Ahn2000, Verhamme2006}, which is composed of a thin \hi\ shell with a single constant radial velocity and a central \lya source. However, due to the symmetry of the shell model, the integrated \lya spectrum is always unpolarized. If the scattering medium is not symmetric, such as a bipolar wind or ellipsoidal halo, the integrated \lya can be polarized \citep{Dijkstra2008,Eide2018}. Thus, to explore polarized \lya, we adopt a new asymmetric model using the radiative transfer code {\it RT-scat} \citep{Chang2023, Chang2024}. We describe the geometry of our wind model and the resulting \lya spectrum in Section~\ref{sec:RT_spec} and polarization in Section~\ref{sec:RT_polarization}. We extensively compare observations and models in Section~\ref{sec:comparison_obsmod}.

\subsection{\lya spectrum in the wind model}\label{sec:RT_spec}
Our new model is composed of a bipolar wind and a point \lya source. 
We show a schematic illustration of the model in Fig.~\ref{fig:geometry}. The wind model is characterized by four main parameters: the \hi\ column density, \NHI, the expansion velocity, \vexp, the half opening angle of the bipolar wind, \openangle, and the width of intrinsic \lya, \sigsrc. The bipolar wind outflows expand radially with a constant velocity \vexp, similar to the shell model. 
The wind's temperature is fixed at 10$^4$ K. The wind's \hi\ number density is constant, and its inner radius is fixed to 10\% of the outer radius. We adopt this geometry as the bipolar wind is intended to represent the outflows outside of the galaxy. 
In our model, we assumed that there is no \hi\ outside the wind outflows, although some \hi\ from inflowing gas or satellites might be present. Given that different \hi\ properties can have opposite effects on the resulting polarization and that we do not have any evidence to constrain these properties, we excluded the presence of additional \hi\ outside the wind outflows, leaving it for future spatially resolved studies (see the discussion in Section~\ref{subsec:future}). 

The central source emits \lya with a Gaussian profile with a width of \sigsrc. This assumption is chosen as it represents the most general case, and is supported by the observational evidence that the galaxy is dust poor (both from the study of optical lines, the blue UV-continuum $\beta$ slope, and the low $E(B-V)$ reddening in \citealt{Iani2021} and from the non detection of dust continuum from ALMA observations in \citealt{Zanella2024}), that could suppress or modify the \lya shape \citep{Laursen2009}. Additionally, we consider the angle of the line of sight \losangle, the azimuthal angle from the $+z$-axis, as the bipolar wind is symmetric about the $z$-axis. In the simulations, we consider $10^6$ photons and extract the escaping \lya spectrum for various \losangle. 

In our new model we assumed a simplified geometry as it allows us to focus on the physical processes that originate polarization. This approach is analogous to that commonly employed to analyze the spectra, where the shell model is usually adopted as the standard model and, even if it does not reflect reality and can be affected by many degeneracies \citep[e.g.,][]{Gronke20162, LiGronke2022}, it can help us to decrypt the information in the \lya line. Currently, there is not a similar standard in the joint study of polarization and spectra together, and we decided to adopt the bipolar wind model as it allows us to explore a large variety of scenarios in a well-motivated physical frame. 

Fig.~\ref{fig:spec_RT} shows the observed $I(\lambda)$ spectrum around the \lya line, the simulated spectrum from the shell model, and the simulated spectra of the wind model. \cite{Iani2021} estimated the physical properties of the shell model (\NHI $\sim 10^{20} \unitNHI$, \vexp $\sim 200 \kms$, and $\sigsrc \sim 100 \kms$), that we adopted as the starting point of our wind model. We found that all the simulated spectra of the wind model for different \openangle values (e.g., we show $\openangle=30$\degree, 60\degree, and 75\degree\ in the left panel of Fig.~\ref{fig:spec_RT}) do not match the observed spectrum, especially near the red peak of the \lya. This is because scattered photons escape outside the outflow opening angle, in the equatorial direction, unlike in the shell model.

To address this discrepancy, we assumed a broader intrinsic \lya (\sigsrc = 150 \kms) to better reproduce the observations. This wider intrinsic spectrum can stem from radiative transfer effects within the inner ISM \citep[e.g.,][]{Gronke2018}.
We explored the effect of this broadening in Appendix~\ref{sec:line_broadening}, finding that the intrinsic \lya profile can broaden from 50 \kms to 150 \kms due to radiative transfer effects occurring within the inner ISM before the photons penetrate into the wind. In addition, we showed in Appendix~\ref{sec:random_motion} that a higher random motion of the bipolar wind cannot reproduce the observed spectrum. In conclusion, a broader intrinsic \lya is required for our modeling with asymmetric geometry.

With this assumption, simulated spectra for \openangle = 60\degree\ and 75\degree\ are similar to those of the previous shell modeling and to the observations, as can be seen in the middle panel of Fig.~\ref{fig:spec_RT}.

\begin{figure*}
    \centering
    \includegraphics[width=\textwidth]{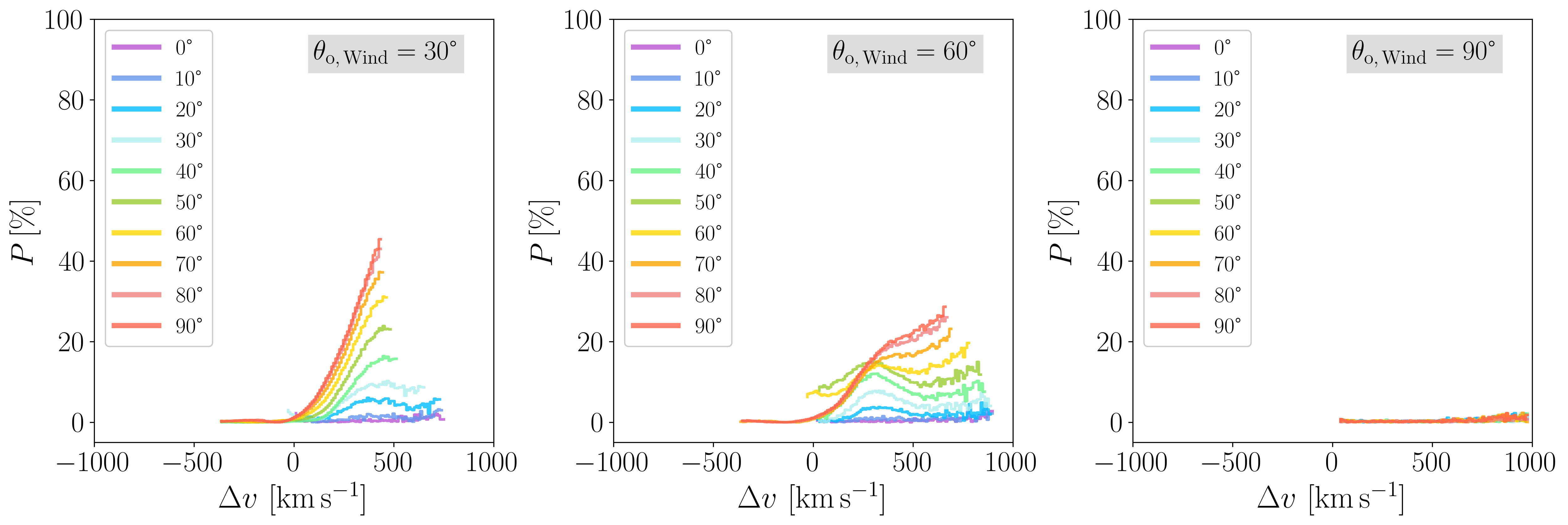}
    \caption{The degree of polarization $P$ for models with \losangle from 0\degree\ to 90\degree, with steps of 10\degree. Given its low significance, we do not display the degree of polarization when the flux in the simulated intensity spectrum is smaller than 5\% of the peak flux of the \lya line.
    The left, middle, and right panels show the results for \openangle = 30\degree, 60\degree, and 90\degree, respectively.
    The overall degree of polarization increases with increasing \losangle and decreasing \openangle.
    }                        
    \label{fig:pol_RT}
\end{figure*}

The right panel of Fig.~\ref{fig:spec_RT} shows the simulated spectra for three \losangle at \openangle = 60\degree. The spectra at \losangle = 0\degree\ and 40\degree\ match the observed spectrum well. On the contrary, the spectrum at \losangle = 90\degree\ does not resemble the observed spectrum and has enhanced flux in the vicinity of the systemic velocity. In the case of a large \losangle, as this one, intrinsic photons are observed directly, without scattering, resulting in a central peak and enhanced red wing in the simulated spectrum. In summary, to reproduce the red peak and the suppressed blue peak of the observed \lya, we need a larger \sigsrc = 150 \kms and \losangle to be smaller than \openangle. The larger \sigsrc allows \lya photons to be emitted over a wider velocity range, broadening the red peak, while the smaller \losangle ensures that most intrinsic photons undergo scattering within the wind cone, suppressing the blue \lya peak emission.

\subsection{Polarization of \lya in the wind model}\label{sec:RT_polarization}

\citet{Dijkstra2008} showed that different models, that assume different production mechanisms of the \lya photons and different geometries of the gas surrounding the emitting source, give rise to different polarization levels. 
From an observational point of view, we detect the signal from an ensemble of photons and not from individual ones, hence the main factor influencing the detected polarization is the geometry of the scattering medium and the presence of a preferential polarization direction, as the signal from highly polarized individual photons may result in an average not polarized signal if their polarization angles are not aligned. Thus,
to understand the polarization behavior of scattered \lya, two fundamental mechanisms to develop polarization are important.

First, the polarization of the integrated \lya strongly depends on the symmetry of the scattering geometry \citep{Eide2018}. Thus, in Fig.~\ref{fig:pol_RT}, at \losangle = 0\degree, the total degree of polarization $P$ becomes zero due to symmetry, regardless of \openangle. Similarly, in the right panel, the total $P$ at \openangle = 90\degree\ is zero for any line of sight. The overall $P$ at \losangle  = 0\degree\ and those at \openangle = 90\degree\ are approximately zero. Additionally, as \openangle increases, the simulated \lya halo becomes more symmetric, leading to a decrease in $P$ (panels from left to right in Fig.~\ref{fig:pol_RT}).

Second, the degree of polarization $P$ of scattered photons increases when the scattering angle, defined as the angle between the incident and scattered directions, approaches 90\degree\ \citep[e.g.,][]{Chandrasekhar60,Chang17,Seon22}. Scattering at angles close to 0\degree\ (forward scattering) or 180\degree\ (backward scattering) causes the $P$ of the scattered photon to be identical to that of the incident photon. Based on this mechanism, when \losangle is close to 90\degree, the fraction of photons undergoing perpendicular scattering increases, leading to a higher $P$. This behavior is evident in the left and middle panels of Fig.~\ref{fig:pol_RT}, where the overall $P$ increases with increasing \losangle. 
As a result, $P$ decreases with increasing \openangle and decreasing \losangle, making the information on geometrical properties of the \hi\ medium imprinted in the polarization of the \lya line. 

Another key factor determining the degree of polarization in symmetric geometries is the \hi\ column density. When \lya emission is polarized and spatially diffused through scatterings, its polarization strongly depends on \NHI. However, this dependence is non-monotonic. \citet{Dijkstra2008} found that the polarization at $\NHI = 10^{20} \unitNHI$ is weaker than at $\NHI = 10^{19} \unitNHI$, due to a higher number of scatterings. In other words, the polarization increases as \NHI decreases. \\
Conversely, in the lower \NHI regime ($< 10^{18} \unitNHI$), the polarization degree decreases with decreasing \NHI due to the dominance of different types of \lya scatterings \citep{Seon22,Chang2023}. Core (resonance) scatterings, which occur at the line center in the rest frame of the \hi\ atom, generally produce weaker polarization compared to wing (Rayleigh) scatterings that occur far from the line center \citep[see also,][]{Ahn2015}. Consequently, metal resonance lines exhibit lower polarization compared to \lya, as discussed further in Section~\ref{sec:metal_polarization}. \\
However, as discussed above, symmetric geometries inherently result in unpolarized spectropolarimetric data due to their symmetry, making them insufficient for reproducing the observed polarization properties of \lya. Thus, in this work we focus on modeling observed data with asymmetric \hi\ geometries.

\subsection{Comparing observations and models}
\label{sec:comparison_obsmod}
We compared our observations of Abell~2895a with the simulated results by using both their total intensity spectra $I(\lambda)$, and the polarization fraction of the \lya, \plya. To properly compare the $I(\lambda)$, we convolved those from simulations with a Gaussian function, to take into account the instrumental spectral resolution of $\sim 3.6 \,  \AA$ ($R \sim 1500$).
We adopted the following physical properties of the wind model (also summarized in Table~\ref{tab:params}): \NHI = $10^{20} \unitNHI$, $\vexp = 200 \kms$, and $\sigsrc = 150 \kms$.
We considered a large variety of geometries for the modeled biconical outflows, varying the wind opening angles \openangle from 0\degree\ to 90\degree\ with steps of 15\degree\ and the line of sight angles \losangle from 0\degree\ to 90\degree\ with steps of 10\degree.
\begin{table}
\scriptsize
\caption{Values of the parameters describing the bipolar wind model adopted in our simulations.}
    \centering
    \begin{tabular}{ccl}
    \hline
    \hline
    Param. & Adopted values & Description\\
    \hline
    \NHI & $10^{20} \unitNHI$ & \hi\ column density \\
    \vexp & $200 \kms$ & Wind expansion velocity \\
    \sigsrc & $ 150 \kms$ & Width of intrinsic \lya \\
    \openangle & 0\degree, 15\degree, 30\degree, 45\degree, 60\degree, 75\degree, 90\degree & Bipolar wind opening angle \\
    \losangle & 0\degree, 10\degree, 20\degree, 30\degree, 40\degree, 50\degree, 60\degree, 70\degree, 80\degree, 90\degree & Line of sight angle \\
    \bottomrule
    \end{tabular}
    \tablefoot{We fix the values of \NHI, \vexp, and \sigsrc (see the main text), while explored different possibilities for \openangle and \losangle, where $\losangle = 0\degree$ means observing in the direction of the outflow.
    }
    \label{tab:params}
\end{table}
We evaluated the goodness of the matching of their total intensities $I$ by adopting a reduced $\chi^2_\nu$ metric over $N=136$ wavelength elements around the \lya emission line, defined as
\begin{equation}
    \chi^2_\nu = \frac{1}{N} \sum_{\lambda = 1206 \, \AA}^{1225 \, \AA} \left[\frac{ I (\lambda) - I_\mathrm{model}(\lambda) }{\sigma_{I}(\lambda)}\right]^2 \, ,
\end{equation}
where $I (\lambda)$, $\sigma_{I} (\lambda)$, and $I_\mathrm{model} (\lambda)$ are, respectively, the observed, its uncertainty, and the modeled normalized total intensity spectra. Thus, models with lower $\chi^2_\nu$ values were preferred, as they indicate that the observed and modeled spectra are more in agreement. We remark that, given that the flux measurements are correlated between several pixels, the $\chi^2_\nu$ values must not be interpreted in an absolute way, but rather qualitatively. We decided to adopt this metric because it can offer a straightforward and clear visualization of the goodness of the agreement, and allows one to directly compare different models, and evaluate those ruled out by $I(\lambda)$.

\begin{figure*}[!h]
    \centering
    \includegraphics[width=0.8\textwidth]{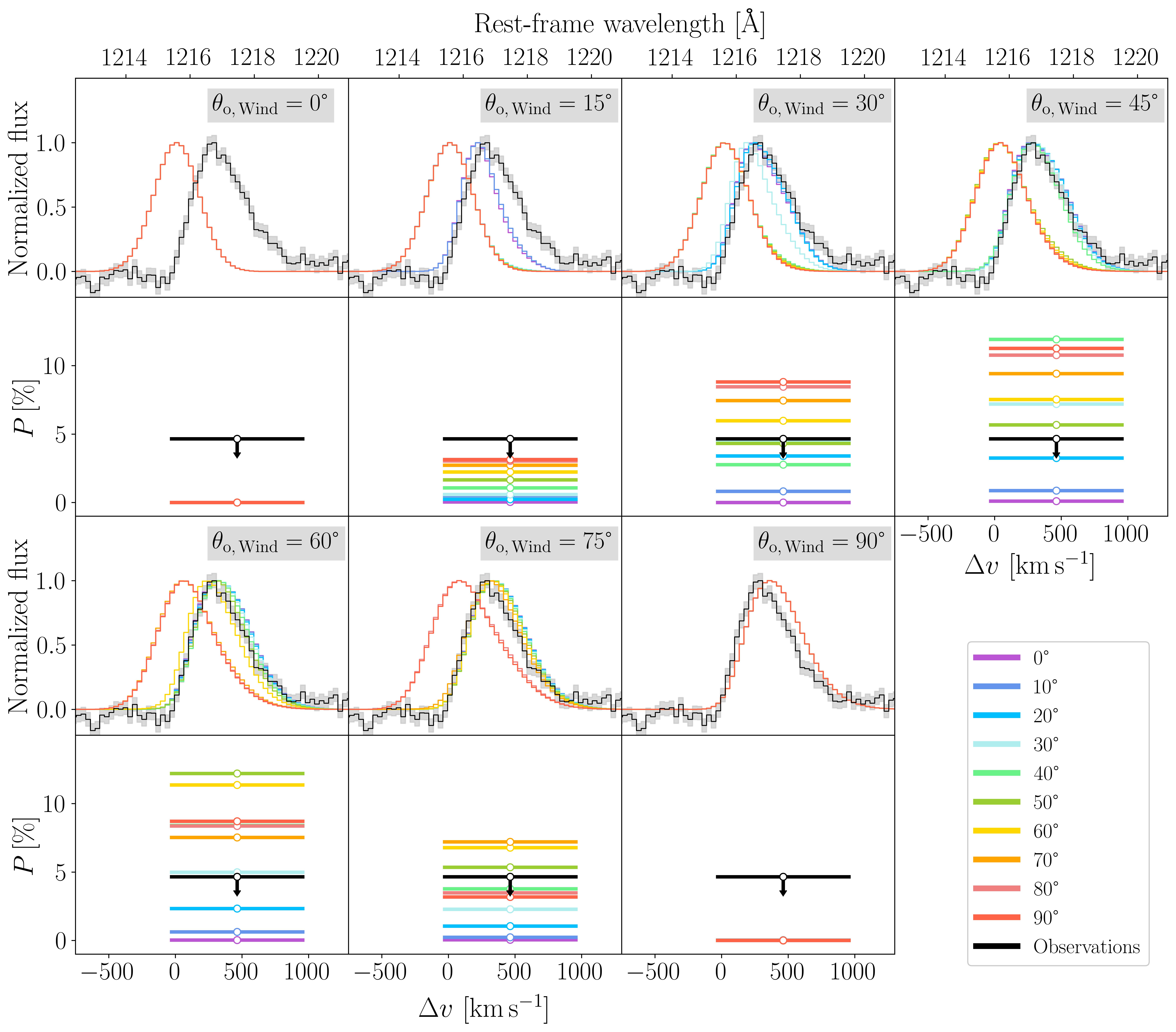}
    \caption{Top and third rows: observed (black, with $1\sigma$ uncertainties in grey) and modeled (with different colors denoting different line-of-sight angles, \losangle, reported in the legend) normalized total intensity spectra $I$, assuming a \hi\ column density \NHI of $10^{20} \unitNHI$, an outflow velocity \vexp of 200 \kms, and a Gaussian width of intrinsic \lya \sigsrc of 150 \kms, in the case of $\openangle=0\degree, 15\degree, 30 \degree, 45 \degree, 60 \degree, 75\degree, 90 \degree$, from left to right, as indicated in the grey labels. Second and bottom rows: Polarization fraction relative to the model described in the corresponding row above. The black open circles represent the observational $1\sigma$ upper limit of $\plya=4.6\%$. The uncertainties for the models are smaller than the linewidths. }                        
    \label{fig:intensities}
\end{figure*}

The observed \lya shows an asymmetric profile that is redshifted with a relative velocity $\Delta v=403 \pm 4$~\kms with respect to the systemic redshift \citep{Iani2021}, as it is typical for outflows. 
In Fig.~\ref{fig:intensities}, we compared observed data and simulated results from the wind models for various \losangle and \openangle.
As we discussed in Section.~\ref{sec:RT_spec}, only the models with $\losangle < \openangle$ are able to reproduce these spectral features (see the top panels of Fig.~\ref{fig:intensities}). 
At $\losangle > \openangle$, the peak of the simulated spectrum is centered on $\Delta v = 0 \kms$, due to directly escaping photons from the central source.
However, at $\openangle \leq 15\degree$, even if $\losangle < \openangle$, the spectrum does not reproduce the observed red wing since a small \openangle induces less scattering.

\begin{figure*}[!h]
    \centering
    \includegraphics[width=0.8\textwidth]{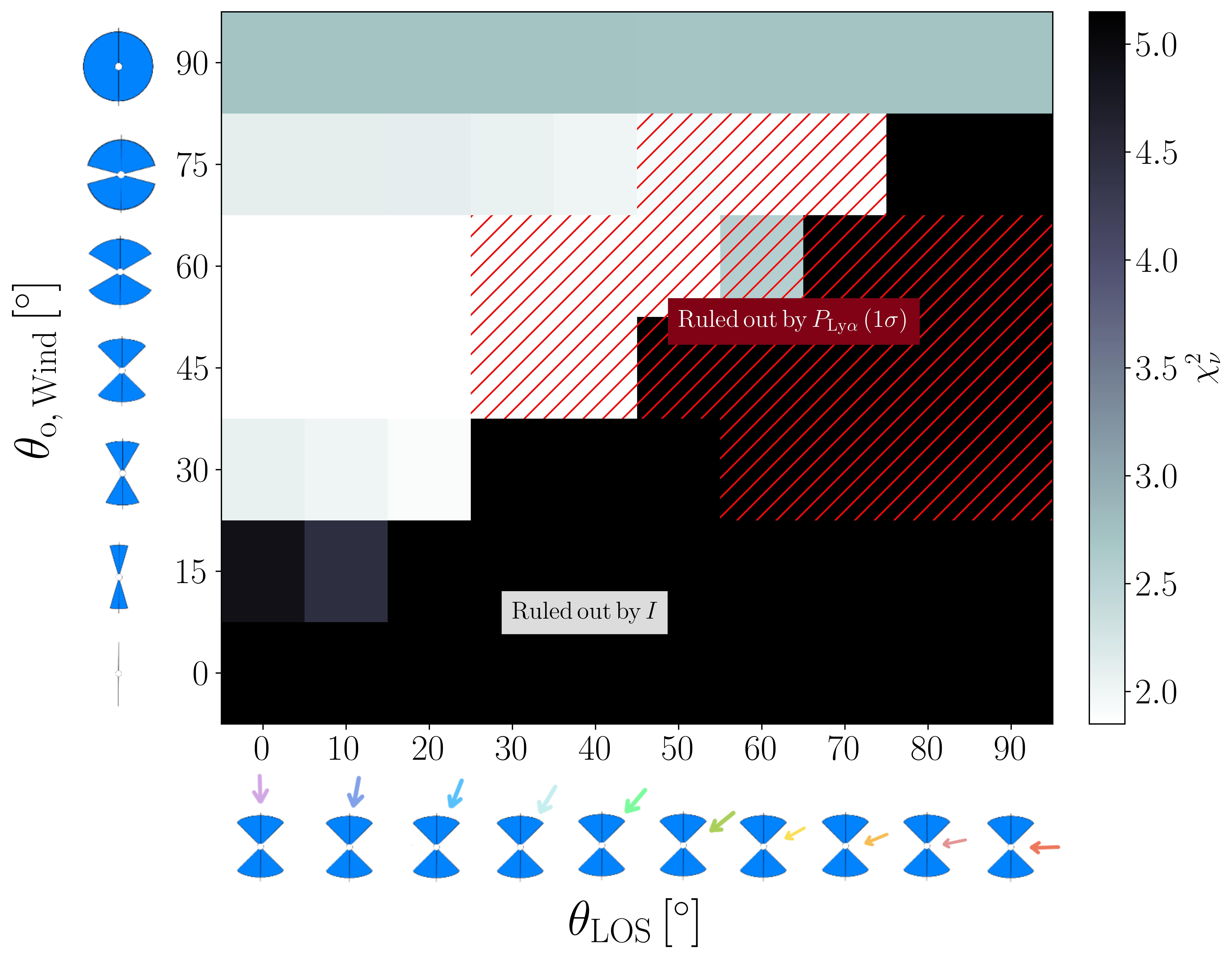}
    \caption{Comparison between observations and wind models at $\NHI = 10^{20} \unitNHI$, $\vexp = 200 \kms$, $\sigsrc =150 \kms$. The sketches along the axes give a visual representation of the models, with increasing \openangle along the $y$-axis and increasing \losangle along the $x$-axis, represented with arrows colored according to Fig.~\ref{fig:intensities}. 
    The grey scale indicates the agreement between the observed and modeled total intensities $I$, evaluated through the $\chi^2_\nu$ value (colorbar on the right), with lighter tones representing better agreement (see the top row of Fig.~\ref{fig:intensities}). Red hatched regions rule out models whose degree of polarization \plya\ is larger than (and thus not consistent with) the observational $1\sigma$ upper limit \plya\ (see the bottom row of Fig.~\ref{fig:intensities}).}
    \label{fig:top_fig}
\end{figure*}

The cases with $\openangle=45\degree$ and $60\degree$ are those better representing the observations, for $\losangle < \openangle$, and have the lowest $\chi^2_\nu$ values. They present an asymmetric profile and can well reproduce the observed red tail, in particular for $\losangle \lesssim 40\degree$. The case with $\openangle=30\degree$ ($75\degree$) has a slightly larger $\chi^2_\nu$ value, because the peak is less (more) redshifted than the observations, but consistent within $1\sigma$. The trend continues to the $\openangle=90\degree$ case, that is the one resembling the expanding ellipsoid geometry of the gas, only marginally consistent with the observed total intensity spectrum. The observed and modeled $I(\lambda)$ spectra for all the models can be seen in the top panels of Fig.~\ref{fig:intensities}, while their $\chi^2_\nu$ values are shown in Fig.~\ref{fig:top_fig} in the greyscale, where lighter tones represent better agreement.  

We binned the degree of polarization from the models with the same binning adopted for the observations, and considered the polarization of the bin including the observed \lya peak, from 1215.5~\AA\ to 1219.6~\AA. Given that we only had upper limits on \plya\ from the observations, we considered as consistent those models whose \lya polarization fraction, within 1$\sigma$, is lower than the observational upper limit. We show the \plya\ from the models and from the observations in the bottom row of Fig.~\ref{fig:intensities}.
As discussed in Section~\ref{sec:RT_polarization}, the low \plya\ value suggests that the observed system is fairly symmetric, and thus the perfectly symmetric $\openangle=0\degree$ and 90\degree\ cases with $\plya \simeq 0$ are fully consistent with observations. The case with $\openangle=15\degree$ ($30\degree$) presents a \plya\ value that increases with \losangle, always consistent (consistent for $\losangle<60\degree$) with the observational constraint. In the cases with $\openangle=45\degree$, $60\degree$, and $75\degree$, the binned \plya\ values do not increase with increasing \losangle, due to the non-vanishing polarization close to the systemic redshift ($\Delta v \approx 0$), that can be seen also in the center panel of Fig.~\ref{fig:pol_RT} (in particular for the $\losangle =50\degree$ and $60\degree$ curves). It results in larger degrees of polarization, that can exceed the observational upper limit for the models with $\openangle \sim \losangle$. The models ruled out by the observational constraints (at the $1\sigma$ level) on the polarization are in hatched red in Fig.~\ref{fig:top_fig}. We highlight that the $I(\lambda)$ and \plya\ constraints rule out complementary regions in the \losangle-\openangle plane, proving the effectiveness of combining these different tracers to investigate the geometry of the scattering \hi\ gas around star-forming galaxies at high redshift and the mechanism of production of the \lya photons.

\section{Discussion} \label{sec:discussion}
\subsection{Origin mechanisms of the \lya} \label{sec:disc:origin}
As introduced in Section~\ref{sec:RT_polarization}, there are two main mechanisms that originate the \lya photons that we observe from distant sources. The first, usually referred to as ``in situ'', includes both the recombination (in photo-ionized gas) and the collisional excitation scenarios.  
Recombination happens when an electron is captured by a proton, resulting in a hydrogen atom in an excited state that can eventually decade to the ground state, emitting \lya \citep{Haiman2001, Cantalupo2005, Arrigoni-Battaia2019}. The probability of emitting \lya mainly depends on the temperature and density of the medium and, in the usually adopted B-case recombination regime, the probability is as large as approximately 68\% at $T=10^4$~K \citep[e.g.,][]{Dijkstra2014}, also explaining why the \lya is the intrinsically stronger emission line in star-forming galaxies. The collisional process involves instead an electron and a hydrogen atom, that is left in an excited state after the close encounter. This process converts
thermal energy of the electrons into radiation, and is thus also called ``\lya emission by cooling radiation''. The second mechanism is related to the scattering, as the \lya photons that are created by a central source (e.g., a star-forming galaxy or AGN), can then escape the production site after numerous scatters in the surrounding \hi\ cloud \citep{Hayes2011, Beck2016}, causing them to propagate far from their production site.
Likely these two mechanisms act simultanesouly \citep[e.g.,][]{Kim2020}. Polarization can be detected in the photoionization case, if enough neutral hydrogen is present in extended regions, producing a considerable scattering probability. 
Moreover, after many scatterings, the emergent \lya line can be depolarized, especially in environments where the medium is isotropic.
We can thus conclude that, if the \lya is polarized, there is photon scattering, while if the \lya is not (or low) polarized, there is no scattering, the gas has a specific geometry, or there are many scatterings.


The \lya intensity spectrum of Abell~2895a requires a scattering contribution to explain the asymmetric profile and the redshifted peak of the line with respect to the systemic redshift. On the other hand, the low polarization upper limit, \plya, is also consistent with photoionization only and no scatter, or with scatter in a fairly symmetric medium that, given the lack of a preferential scattering direction, would result in a measured low polarization signal, even if the individual photons might be highly polarized. For example, as described in Section~\ref{sec:comparison_obsmod} for Abell~2895a, centrally emitted photons that experience scattering can give rise to an integrated $\plya \sim 0$, consistent mostly with the large \openangle and low \losangle cases, that are the most symmetric ones. These preferred values of \openangle and \losangle are also consistent with the observed displacement of $1.2 \pm 0.2$~kpc \citep[on the source plane,][]{Iani2021} between the UV stellar continuum and the peak of the \lya, if we assume that the \lya photons are emitted by the stars and that only one side of the wind is illuminated (it can be explained with different scenarios, e.g., an initial transient feature before breaking through the disk, an enhanced dust obscuration of the far wind, or an effect of tidal stripping).

Deep spatially resolved polarization observations, that we discuss in the next section, are needed to reconstruct in detail the production site of \lya photons and the geometry of the scattering medium, allowing us to disentangle between the different possible production mechanisms of the \lya.

\subsection{Spatially resolved observations and future perspectives}
\label{subsec:future}
Scattering of \lya photons through a geometrically asymmetric medium give rise to polarized signal \citep{Angel1969, Lee1998}, detectable also without resolving the system \citep{Eide2018}, as we described in Sections~\ref{sec:analysis3} and \ref{sec:radiative_transfer} for Abell~2895a. However, as discussed in Section~\ref{sec:disc:origin}, this leaves the question about the origin of the \lya photons (and thus the region from where they are emitted) and the geometry and orientation of the scattering medium. This degeneracy can be broken by learning the geometry of the galaxy from other independent probes, or by using  spatially resolved and deeper spectropolarimetric observations. Spatially resolved observations can put constraints on the polarization over different regions that we can compare with models predicting different degrees of polarization ranging, for example in a biconical wind geometry, from $\plya \sim 0\%$ in the center to \plya\ up to $80\%$ in the outflows \citep{Eide2018}. Additionally, deeper observations can help putting tighter constrains to the measured polarization levels, and achieving observational $S/N$ high enough to measure also the polarization angle, that is predicted to significantly change between different models, giving hints on, for example, the direction of the outflow in a biconical wind geometry \citep{Eide2018}. As described in Section~\ref{sec:analysis3} and in Appendix \ref{app_b}, we tentatively measured the polarization fraction by considering only half of the \lya emission, with the goal of including these information in the comparison with the models. But, due to the low $S/N$ and angular resolution of our observations and the moderate tangential stretch from lensing, we could only measure \lya polarization upper limits of approximately 7.5\% at 1$\sigma$.

Currently, Abell~2895a is the only distant star-forming galaxy with \lya spectropolarimetric observations. The main reasons for the lack of such observations are the typical low luminosity of high-$z$ star-forming galaxies and the fact that the light beam has to be split in two components (its $o$ and $e$ components, with orthogonal polarizations) and rotated, significantly increasing the observational exposure time necessary to reach sufficient $S/N$ values, even with the most advanced spectropolarimeters mounted on cutting-edge telescopes.
This issue can be mitigated by observing galaxies at $z\sim 2-4$ (with redshifted \lya line included in the instrument's wavelength coverage) that are strongly lensed (i.e., magnified and distorted) with a large magnification factor ($\mu \gtrsim 10$), that would sensibly lower the needed exposure time and increase the spatial resolution. The observation of the most magnified and distorted sources lying across the lensing critical lines would allow us to zoom-in and study the spatial variation of the polarization across the \lya emission on sub-kpc scales and put more stringent limits to the origin of the \lya emission and the scattering \hi\ geometry. Additionally, extending such observations to a small sample of galaxies will also allow us to put the results obtained for Abell~2895a in a broader context.

\subsection{Polarization of other resonant lines}
\label{sec:metal_polarization}

In principle, the study presented in this paper can be executed by employing different resonant line emissions than the \lya, like the \mgii, \civ, \ovi, \nv, \siiv\ doublets, commonly used in astrophysics \citep[e.g.,][]{Prochaska2011, Hayes2016, Henry2018, Berg2019, Katz2022, Dutta2023}. The scattering processes of such resonance doublets are very similar to those of the \lya line, because they present similar atomic properties with one electron in the outer orbit and have the same atomic structure composed of two transitions, called ``K'' ($S_{1/2} - P_{3/2}$) and ``H'' ($S_{1/2} - P_{1/2}$). 
Due to their resonance nature, metal resonance doublets are also spatially extended and polarized via scattering.
The inclusion of such lines would allow us to expand the usable redshift range (atomic line center wavelengths from $\sim 1032 \,\AA$ for the \ovi\ to $\sim 2800 \,\AA$ for the \mgii) and to explore the multi-phase nature of the CGM, exploring gas temperatures from $10^{3-4}$~K (ionization energies of 13.6~eV for the \lya and 15~eV for the \mgii) to $10^{5-6}$ K (138~eV of the \ovi). 
Moreover, the two K and H transitions, that are not possible to observationally disentangle for the \lya line, because of the small energy difference of $\sim 10^{-4}$~eV $\approx 1.5 \kms$, are resolved for other lines, like the \mgii\ (transition K at 2796.4~\AA\ and H at 2803.5~\AA, with a separation of $760 \kms$) or the \siiv\ (K at 1393.8 \AA\ and H at 1402.8 \AA, with the largest separation of $1926 \kms$ between the above-mentioned resonant lines). 

Recently, radiative transfer models that include the \mgii\ doublet have been developed, also including polarization \citep{Seon23, Chang2024}. In particular, \citet{Chang2024} considered three-dimensional shell, sphere, and clumpy sphere geometries, and studied the joint \lya and \mgii\ escape using \textit{RT-scat}. They reveal that, despite being driven by similar atomic processes, the emerging \mgii\ and \lya spectra are very different, because of the different \mgii\ and \hi\ column densities, with $\NHI \gg { N}_{{\rm Mg}\textsc{ii}}$. 
Moreover, they confirmed a correlation between the escape of LyC radiation and the \mgii\ double ratio, the ratio of the K and H lines \citep[also see,][]{Henry2018, Chisholm2020, Izotov2022, Katz2022, Xu2023}.
They also found that the \mgii\ degree of polarization decreases with increasing column density ${ N}_{{\rm Mg}\textsc{ii}}$, because the multiple scatterings without a preferential direction decrease the resulting polarization. This results in a low ($<5\%)$ degree of polarization in the center, growing up to 10\% (25\%) with ${ N}_{{\rm Mg}\textsc{ii}} = 10^{14}$~cm$^{-2}$ (${ N}_{{\rm Mg}\textsc{ii}} = 10^{13}$~cm$^{-2}$). 
Consequently, they suggested that the \mgii\ polarization can be used to estimate the doublet ratio of the extended halo.
Unfortunately, the \mgii\ emission is much fainter than the \lya, making it challenging to resolve and measure its degree of polarization, and requiring extremely large amount of observational time, especially at high redshift. 
Again, strongly lensed candidates would allow us to obtain spatially extended and spectropolarimetric observations and to pursue this kind of studies.

\section{Summary and conclusions}
\label{sec:summary}
In this paper we put novel constraints on the geometry of the \hi\ region surrounding a clumpy star-forming galaxy at $z \approx 3.4$, Abell~2895a \citep{Livermore2015, Iani2021, Zanella2024}, strongly lensed into three multiple images by the cluster of galaxies Abell~2895. We made use of new VLT/FORS2 observations taken with the Polarimetric Multi Object Spectroscopy (PMOS) mode to extract the spectra relative to the ordinary ($o$) and extraordinary ($e$) beams in four different half-wave plate angles, from an aperture including the M1 and M2 multiple images of Abell~2895a. We combined the different spectra to measure the total intensity 1D spectrum, $I(\lambda)$. We focused on the \lya emission line, and we computed the Stokes parameters $q$ and $u$, used to measure the polarization fraction \plya, corrected both for the positive bias of polarization and for the dilution effect due to the contamination of the angularly close, unpolarized, BCG of the Abell~2895 cluster. 
We obtained \plya\ upper limits of 4.6\%, 5.8\%, and 6.5\% at the $1\sigma$, $2\sigma$, and $3\sigma$ level. We showed that polarization constraints are complementary to those usually achieved by using only the total \lya intensity and line profile, demonstrating the effectiveness of this novel technique.

To interpret the observational constraints, we developed a \lya radiative transfer model 
including a bipolar wind geometry (characterized by the half opening angle \openangle, the radial expansion velocity \vexp, and the \hi\ column density \NHI parameters) and a central source that emits \lya photons with a Gaussian profile with a width of \sigsrc. We assumed \NHI $\sim 10^{20} \unitNHI$, \vexp $\sim 200 \kms$, and $\sigsrc \sim 150 \kms$ and tried different \openangle values (from 0\degree\ to 90\degree\ with steps of 15\degree) and simulated different line of sight angles (\losangle from 0\degree\ to 90\degree\ with steps of 10\degree). 

In order to reproduce the spectral profile and shift with respect to the systemic velocity of the \lya line, we needed \lya photons that are created in the inner regions and that undergo several scattering events before escaping, and $\losangle~<~\openangle$ and $\openangle > 15\degree$. 
Models with \openangle of 45\degree, 60\degree, or 75\degree\ with \losangle of $30\degree-60\degree$ can well reproduce the \lya spectral profile, but the combination of \openangle and \losangle\ resulting in an asymmetric geometry and the scattering due to the larger \openangle, make the predicted polarization much larger than the low observed \plya, that is consistent with no scattering or with scattering in a fairly symmetric medium, and thus are excluded. 

Summarizing, the models that satisfy both the \lya spectral profile and polarization requirements are those with \openangle $\sim$ 30\degree\ for \losangle $\leq 20\degree$, \openangle $\sim$ 45\degree\ for \losangle $\leq 20\degree$, \openangle $\sim$ 60\degree\ for $\losangle~\leq~20\degree$, \openangle $\sim$ 75\degree\ for $\losangle \leq 40\degree$, and \openangle $\sim$ 90\degree\ for any \losangle, where $\losangle = 0\degree$ means observing in the direction of the outflow. These results will pave the way to future spatially resolved spectropolarimetric observations, needed to discriminate between the different production mechanisms of the \lya and the geometry of the scattering medium.

\begin{acknowledgements}
We are grateful to the referee, Matthew Hayes, for the insightful comments and constructive suggestions, which have significantly enhanced the quality of this paper.
This research is based on observations collected at the European Organisation for Astronomical Research in the Southern Hemisphere under ESO programme 108.2260, It is also based on observations made with the NASA/ESA Hubble Space Telescope, and obtained from the Hubble Legacy Archive, which is a collaboration between the Space Telescope Science Institute (STScI/NASA), the Space Telescope European Coordinating Facility (ST-ECF/ESA) and the Canadian Astronomy Data Centre (CADC/NRC/CSA). \\
The research activities described in this paper have been co-funded by the European Union – NextGeneration EU within PRIN 2022 project n.20229YBSAN - Globular clusters in cosmological simulations and in lensed fields: from their birth to the present epoch.
The authors thank Sabine Moehler for the helpful discussions regarding the FORS2 PMOS data reduction. 
A.B. and A.Z. acknowledge support from the INAF minigrant 1.05.23.04.01 ``Clumps at cosmological distance: revealing their formation, nature, and evolution''.
M.G. thanks the Max Planck Society for support through the Max Planck Research Group. 
E.I. acknowledges funding from the Netherlands Research School for Astronomy (NOVA). 
We acknowledge the use of the \texttt{numpy} \citep{numpy}, \texttt{matplotlib} \citep{matplotlib}, \texttt{astropy} \citep{astropy}, and \texttt{pandas} \citep{pandas} packages.
\end{acknowledgements}

\begin{appendix}
\section{\lya polarization measured in the three-bin case}
\label{app_a}
\begin{figure}
    \centering
    \includegraphics[width=\columnwidth]{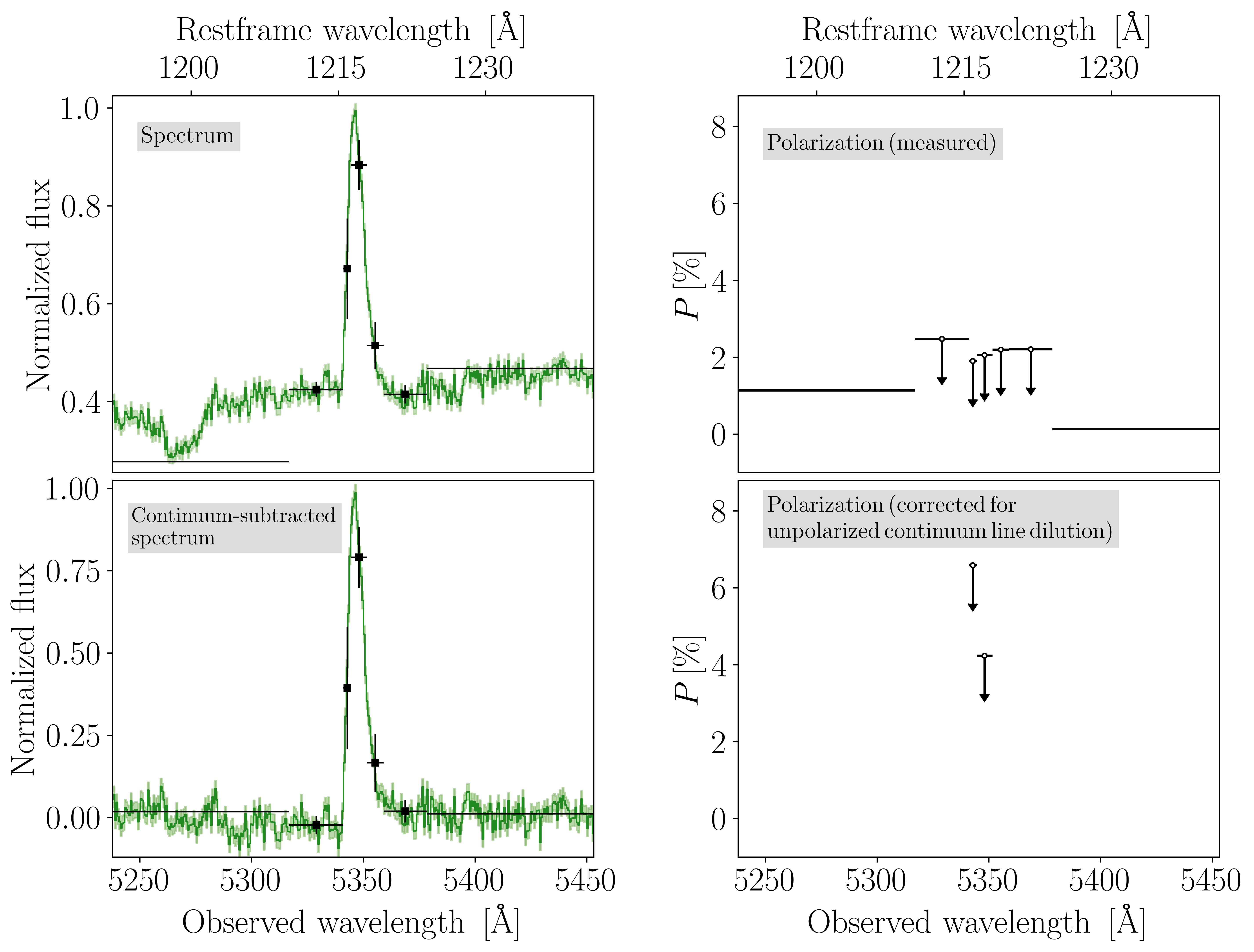}
    \caption{Left panels: total intensity spectra (green solid line) and 1$\sigma$ uncertainties (light green shaded region) in the spectral region around the \lya line, similar to those presented in Fig.~\ref{fig:pre_post_dil_corr}, but in the three-bin case. The top (bottom) panel shows the normalized spectrum before (after) the subtraction of the BCG contribution. The black filled squares, with 1$\sigma$ uncertainties, represent the binned data. Right panels: polarization ($P$) measurements obtained before (top) and after (bottom) the dilution correction described in Eq.~\ref{eq:dilution}. The black open circles represent the $1\sigma$ upper limits obtained by applying the correction of \citet{Simmons1985}. In the bottom right panel, only two datapoints are visible, as the others are outside the plotted range (upper limits for $P$ from 13 to 95\%).}
    \label{fig:pre_post_dil_corr_appendix}
\end{figure}

We show here the results obtained by dividing the \lya line over three bins, one including the blue tail (1215.5-1216.3~\AA), a central one including the peak (1216.3-1217.9~\AA) and one including the red tail (1217.9-1219.6~\AA). The binning is completed by sampling the blue and red continua with two bins: one narrower and closer to the line and one broader, from 1210.0~\AA\ to 1215.5~\AA, and from 1037~\AA\ to 1210~\AA\ for the blue, respectively, and from 1219.6~\AA\ to 1224.0~\AA, and from 1224~\AA\ to 1260~\AA\ for the blue, respectively. The bins are shown in the left panels of Fig.~\ref{fig:pre_post_dil_corr_appendix}. For the blue tail, central, and red tail bins, we measure $1\sigma$ upper limits on \plya of 6.6\%, 4.2\%, and 13.1\%, respectively. Due to the large dilution factor, the polarization degree is barely (not) constrained for the narrow (broad) continua bins.

\section{Tentative spatially resolved \lya polarization measurement}
\label{app_b}
\begin{figure}
    \centering
    \includegraphics[width=\columnwidth]{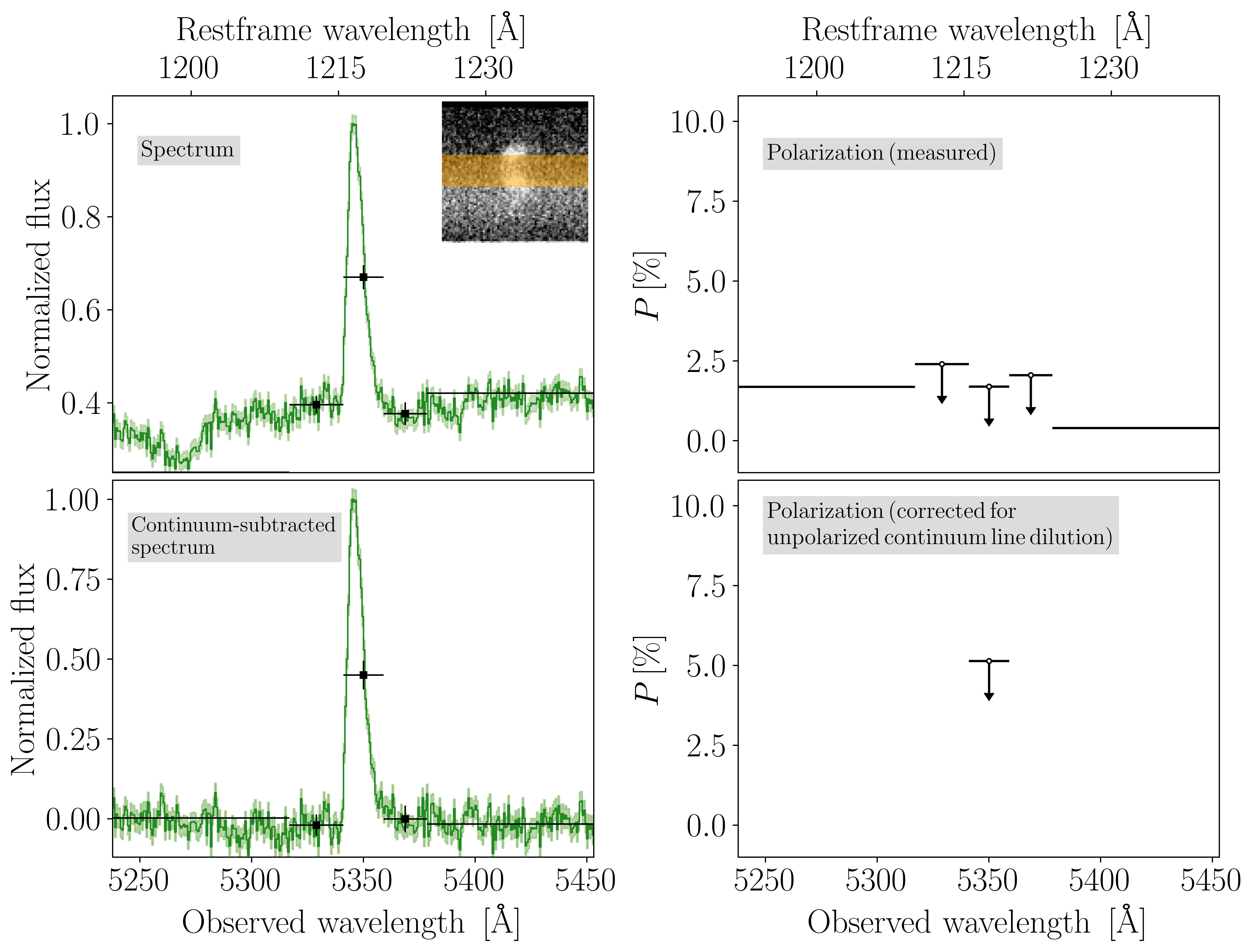}
    \caption{Left panels: total intensity spectra (green solid line) and 1$\sigma$ uncertainties (light green shaded region) in the spectral region around the \lya line, similar to those presented in Fig.~\ref{fig:pre_post_dil_corr}, but extracted from a region designed to include a spatial half of the \lya blob. The aperture, represented in orange in the inset, has a width of $\approx 4''$ and goes from the M1 \lya peak to the M2 \lya peak. The top (bottom) panel shows the normalized spectrum before (after) the subtraction of the BCG contribution. The black filled squares, with 1$\sigma$ uncertainties, represent the binned data. Right panels: polarization ($P$) measurements obtained before (top) and after (bottom) the dilution correction described in Eq.~\ref{eq:dilution}. The black open circles represent the $1\sigma$ upper limits obtained by applying the correction of \citet{Simmons1985}.}
    \label{fig:app_b_int}
\end{figure}

\begin{figure}
    \centering
    \includegraphics[width=\columnwidth]{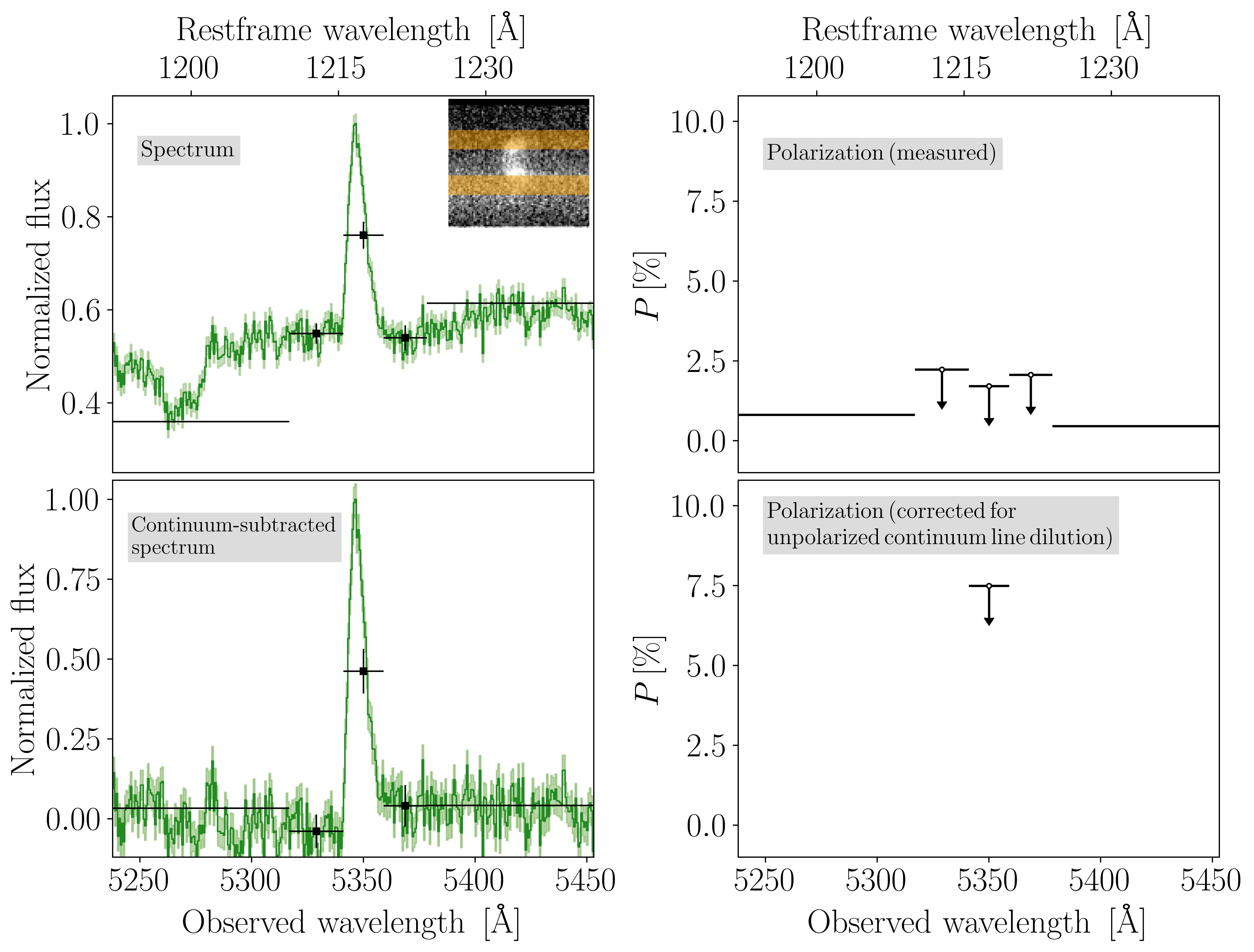}
    \caption{As in Fig.~\ref{fig:app_b_int}, but with the spectra extracted from the other spatial half of the \lya emission. In particular, the extraction aperture, represented in orange in the inset, has a total width of $6''$, divided into two apertures of $3''$, from each \lya peak towards the edge of the slit.}
    \label{fig:app_b_ext}
\end{figure}

We show here the results obtained by dividing the spatial extraction aperture over two different regions, in order to verify whether the low polarization fraction measured might be the result of the cancellation of opposite contributions from different sides. Unfortunately, the $S/N$ is not sufficient to perform such test on the individual M1 and M2 multiple images but, thanks to their mirrored nature, we designed two apertures to include the same spatial half of each \lya emission. In particular, the first aperture has a width of $\approx 4''$ and goes from the M1 \lya peak to the M2 \lya peak. The second aperture has a total width of $6''$, divided into two apertures of $3''$, from each \lya peak towards the edge of the slit. 
Considering the case with a single bin for the \lya line, we obtain upper limits of 5.1\%, 7.1\%, and 8.3\% at the 1$\sigma$, 2$\sigma$, and 3$\sigma$ level for the first aperture, while 7.5\%, 10.2\%, and 11.9\%, respectively, for the second aperture. The results are shown in Fig.~\ref{fig:app_b_int} and \ref{fig:app_b_ext}. The results show that low polarization levels are measured in different spatial regions, and thus that the golbal low polarization is not the result of cancellation of the Stokes vectors from two different sides. Unfortunately, due to the low $S/N$ and spatial resolution, it is not possible to include the spatial information in our analysis and in a more detailed comparison with the model.


\section{\lya Line Broadening}\label{sec:line_broadening}

\begin{figure}
    \includegraphics[width=0.45\textwidth]{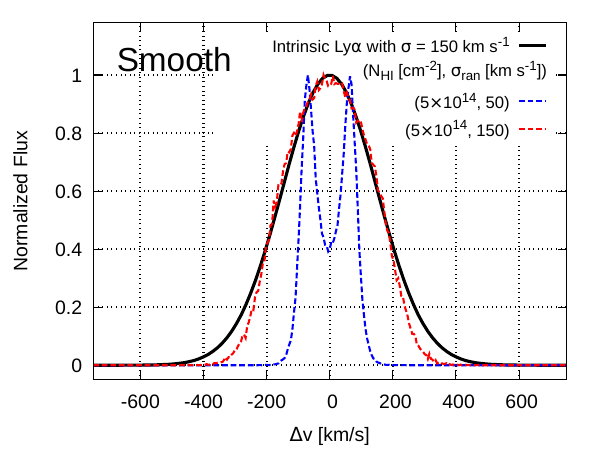}
    \includegraphics[width=0.45\textwidth]{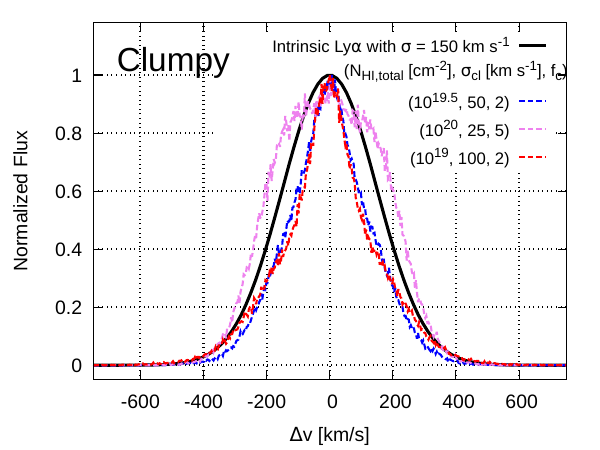}
    \caption{
    Simulated spectra in the spherical geometry considering smooth (top) and clumpy (bottom) media.
    The black solid lines represent Gaussian functions with a width of $\sigma = 150 \kms$, corresponding to the intrinsic \lya profile estimated using the wind model in Section~\ref{sec:RT_spec}.
    }                 
    \label{fig:braod_RT}
\end{figure}

In Section~\ref{sec:RT_spec}, our modeling with the wind geometry required the intrinsic \lya emission with $\sigsrc = 150 \kms$ to reproduce the observed spectrum.
However, this is three times broader than the observed width of $\sim 50$~\kms of the H$\beta$ line \citep{Iani2021}. This is puzzling as the intrinsic \lya width should be similar to the width of the observed Balmer lines. Similar discrepancies have been noted previously in the literature \citep{2016ApJ...820..130Y,2018A&A...616A..60O}.
In this section, we aim to understand the broad intrinsic \lya through the effect of \lya radiative transfer in a small-scale \hi\ gas in ISM or star-forming regions since its resonance nature induces the broadening of the escaping spectrum and can potentially explain the differences between the line widths \citep{Neufeld1990,Gronke2018}.


To test this broadening effect, we assumed a spherical geometry composed of a central \lya point source surrounded by a static spherical \hi\ halo. The line width of the central \lya emission was fixed at 50~\kms, corresponding to the H$\beta$ width. Two types of \hi\ gas distributions were considered: a smooth medium with uniform \hi\ density and a clumpy medium composed of small \hi\ clumps in an empty inter-clump medium, where the clump size is 100 times smaller than the halo size. The smooth medium is characterized by the \hi\ column density, \NHI, and the random motion, $\sigma_{\rm ran}$. The clumpy medium is described by the total \hi\ column density, $N_{{\rm H\textsc{i}, \, total}}$, the clump random motion, $\sigma_{\rm cl}$, and the covering factor, $f_{\rm c}$, which represents the mean number of clumps along the line of sight from the central source \citep[see, e.g.,][]{Hansen06}. The total column density $N_{{\rm H\textsc{i}, \, total}}$ is $f_{\rm c} N_{{\rm H\textsc{i}, \, cl}}$, where $N_{{\rm H\textsc{i}, \, cl}}$ is the \hi\ column density of a single clump. Details of this geometric setup are described in \citet{Chang2024}. Fig.~\ref{fig:braod_RT} illustrates simulated spectra (dotted lines) compared to a Gaussian profile with $\sigsrc = 150 \kms$ (black solid lines).

For the smooth medium, we fixed $\sigma_{\rm ran}$ at 50 \kms, corresponding to the intrinsic \lya width. However, in the top panel of Fig.~\ref{fig:braod_RT}, the simulated spectrum exhibits a double-peak profile that does not match the Gaussian shape. This outcome arises because optically thick \hi\ medium at the \lya line center generally produces double-peaked profiles \citep{Neufeld1990}. To resolve this, we decreased \NHI for optically thin gas at the line center and increase $\sigma_{\rm ran}$ to enhance spectral broadening. When $\sigma_{\rm ran}$ reaches 150 \kms, the simulated spectrum closely matches the Gaussian profile.

In the clumpy medium, we considered small values of $f_{\rm c}$ since low covering factors enable single-peak profiles via surface scattering (\citealp{Neufeld91}; see also \citealp{Hansen06, Chang2023}) when $f_{\rm c}$ is below the critical covering factor \citep[][]{gronke_clumpy16,Gronke2017}. In the bottom panel of Fig.~\ref{fig:braod_RT}, the simulated spectra successfully reproduce the Gaussian profile with $\sigsrc = 150 \kms$. 

In summary, the intrinsic \lya profile can broaden easily from 50 \kms to 150 \kms due to radiative transfer effects occurring within the inner ISM before the photons penetrate into the wind. We can simulate this effect within the context of a clumpy medium in which a relatively small number of clumps ($f_{\rm c}\lesssim f_{\rm c,crit}$) can lead to a Gaussian line broadening, or in a smooth medium using a large turbulent motion $\sigma_{\rm ran}$ and a low \NHI. These findings underscore the distinct mechanisms by which \lya line formation is influenced by the physical properties of small-scale cold gas.

\section{Impact of turbulent motion within a bipolar wind on the formation of \lya}
\label{sec:random_motion}

\begin{figure}
    \centering
    \includegraphics[width=0.45\textwidth]{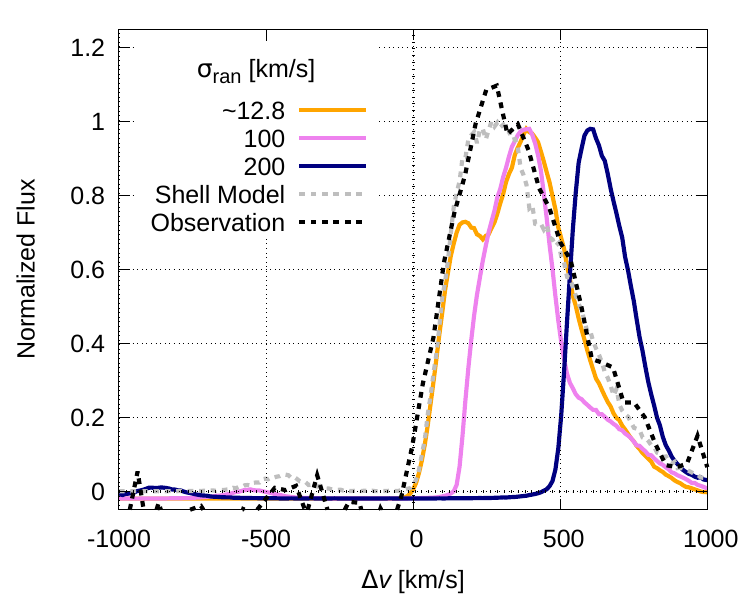}
    \caption{
    Simulated \lya spectra for various random motions of the \hi\ bipolar wind ($\sigma_{\rm ran} = 12.8$, 100, and 200 \kms) at \openangle = 60\degree\ and \losangle = 0\degree. The intrinsic line width \sigsrc is fixed at 100 \kms, based on the shell model in \cite{Iani2021}. The black and gray dashed lines represent the observed spectrum and the shell model spectrum, respectively. The orange line corresponds to the spectrum at \openangle = 60\degree\ from Fig.~\ref{fig:spec_RT}. The navy and violet lines are spectra for $\sigma_{\rm ran}$ = 100 and 200 \kms, respectively. Higher random motion induces redshifting and deviation from the observed spectrum. The simulated spectra are normalized by setting the same peak height as the spectrum of the previous fit with the shell model.
    }           
    \label{fig:random_motion}
\end{figure}

Galactic winds, as most astrophysical media, are highly turbulent which is imprinted, for instance, in the large random velocities seen in both simulations and observations orthogonal to the bulk flow direction \citep[e.g.,][]{Schneider2020, Veilleux2020}. In this section, we want to investigate the effect of this turbulent component on the emergent \lya spectrum.


Fig.~\ref{fig:random_motion} shows \lya spectra for three different random motions of the bipolar wind: $\sigma_{\rm ran} = 12.8$, 100, and 200 \kms.\footnote{Note that while turbulent velocities of $\sim 100$~\kms seem large, the hot component in galactic winds can easily be $\gtrsim 10^7\,$K, thus, the turbulence is still highly subsonic.} The spectrum at $\sigma_{\rm ran} = 12.8 \kms$ corresponds to the model at \openangle = 60\degree\ in the left panel of Fig.~\ref{fig:spec_RT}. As $\sigma_{\rm ran}$ increases, the spectrum becomes more redshifted and deviates from the observed spectrum. This behavior arises because higher random motion induces a more significant peak shift in the \lya line. 

Consequently, accounting for increased random motion within this set of other parameters, the bipolar wind model fails to reproduce the observed spectrum. However, naturally other degeneracies exist which are, however, beyond the scope of this study to explore. Interestingly, the shell-model fit of \citet{Iani2021} produces an effective temperature of $\log(T_{\rm eff}/{\rm K})=5.3\pm 0.2$, that is, assuming a temperature of $T=10^4\,$~K corresponding to a random motion of $\sim 50$~\kms.


\end{appendix}

\bibliographystyle{aa} 
\bibliography{bibliografia} 

\end{document}